\documentclass[a4paper,11pt]{article}
\usepackage{jheppub}
\usepackage{amsmath,amssymb,bm,mathrsfs}
\usepackage{mathtools,mathptmx,array,slashed}
\usepackage{graphicx,graphics,subfigure}
\usepackage{ctable,multirow}
\usepackage{tabularx}
\usepackage[colorlinks=true,linkcolor=blue,citecolor=blue,urlcolor=blue]{hyperref}

\def\be{\begin{equation}}
\def\ee{\end{equation}}
\def\bea{\begin{eqnarray}}
\def\eea{\end{eqnarray}}
\def\ba{\begin{aligned}}
\def\ea{\end{aligned}}
\def\nn{\nonumber}
\def\p{\partial}

\title{Topological classes of thermodynamics of the static multi-charge AdS black holes in gauged supergravities:
novel temperature-dependent thermodynamic topological phase transition}

\author{Di Wu,}
\author{Shuang-Yong Gu,}
\author[1]{Xiao-Dan Zhu\note{Corresponding author.},}
\author[1]{Qing-Quan Jiang}
\author[1]{and Shu-Zheng Yang}

\affiliation{School of Physics and Astronomy, China West Normal University, \\
1 Shida Road, Nanchong, Sichuan 637002, People's Republic of China}

\emailAdd{wdcwnu@163.com, shuangyonggu@gmail.com, zxdcwnu@163.com, qqjiangphys@yeah.net, szyangcwnu@126.com}

\abstract{In this paper, we investigate, in the framework of the topological approach to black hole thermodynamics, using the generalized off-shell Helmholtz free energy, the topological numbers of the static multi-charge AdS black holes in four- and five-dimensional gauged supergravities. We find that the topological number of the static-charged AdS black holes in four-dimensional Kaluza-Klein (K-K) gauged supergravity theory is $W = 0$, while that of the static-charged AdS black holes in four-dimensional gauged $-iX^0X^1$-supergravity and STU gauged supergravity theories, and five-dimensional Einstein-Maxwell-dilaton-axion (EMDA) gauged supergravity and STU gauged supergravity, and five-dimensional static-charged AdS Horowitz-Sen black hole are both $W = 1$. Furthermore, we observe a novel temperature-dependent thermodynamic topological phase transition that can happen in the four-dimensional static-charged AdS black hole in EMDA gauged supergravity theory, the four-dimensional static-charged AdS Horowitz-Sen black hole, and the five-dimensional static-charged AdS black hole in K-K gauged supergravity theory. We believe that the novel temperature-dependent thermodynamic topological phase transition could help us better understand black hole thermodynamics and, further, shed new light on the fundamental nature of gauged supergravity theories.}


\arxivnumber{2402.00106}

\begin{document}

\maketitle

\flushbottom

\section{Introduction}
The discovery of the anti-de Sitter/conformal field theory (AdS/CFT) correspondence \cite{ATMP2-231,PLB428-105,ATMP2-253} has attracted
a great deal of interest in studying the thermodynamic properties of charged AdS black holes in four- and five-dimensional gauged
supergravities \cite{NPB554-237,NPB553-317,PRD72-084028,PRD84-024037,PRL95-161301,hep-th/0504080,PRD72-041901,PRD73-104036,PRL100-121301,
PRD80-044037,PRD80-084009,PRD83-044028,PRD83-121502,PLB707-286,PLB726-404,PLB746-276,PRD101-024057,PRD102-044007,PRD103-044014,JHEP1121031}.
In fact, the establishment of the three widely accepted thermodynamic mass formulas, i.e., the first law of black hole thermodynamics
\cite{PRD7-2333,PRD13-191}, the Bekenstein-Smarr mass formula \cite{PRL30-71}, and the Christodoulou-Ruffini squared-mass formula
\cite{PRL25-1596,PRD4-3552}, is not the only facet of the investigation of black hole thermodynamics.

Recently, topology has received considerable interest and enthusiasm as an important mathematical tool applicable to black hole physics.
There are two important aspects to the topology research underway at present. One area of investigation focuses on the light rings
\cite{PRL119-251102,PRL124-181101,PRD102-064039,PRD103-104031,PRD105-024049,PRD108-104041,2401.05495} corresponding to some black holes,
which could present more support for black hole observation in the future and has been expanded to timelike circular orbits \cite{PRD107-064006,
JCAP0723049}. Another area of investigation focuses on the thermodynamic properties of black holes \cite{PRD105-104003,PRD105-104053,
PLB835-137591,PRD107-046013,PRD107-106009,JHEP0623115,2305.05595,2305.05916,2305.15674,2305.15910,2306.16117,PRD106-064059,PRD107-044026,
PRD107-064015,2212.04341,2302.06201,2304.14988,2309.00224,2312.12784,2402.18791,2403.14730,2404.02526}. Extraordinarily, a new method for
exploring the thermodynamic topological features of black holes has developed, as pointed out in Ref. \cite{PRL129-191101}. This method
treats black hole solutions as topological thermodynamic defects, calculates topological numbers, and then classifies black holes into
three different categories according to their topological numbers. This breakthrough work has provided novel insights into the fundamental
nature of black holes and gravity. The thermodynamic topological procedure presented in Ref. \cite{PRL129-191101} has achieved considerable popularity
because of its broad application and convenience. Therefore, it has been effectively applied to explore the topological numbers corresponding
to various famous black holes \cite{PRD107-064023,JHEP0123102,PRD107-024024,PRD107-084002,PRD107-084053,2303.06814,2303.13105,2304.02889,2306.13286,
2304.05695,2306.05692,2306.11212,EPJC83-365,2306.02324,PRD108-084041,2307.12873,2309.14069,AP458-169486,2310.09602,2310.09907,2310.15182,
2311.04050,2311.11606,2312.04325,2312.06324,2312.13577,2312.12814,PS99-025003,2401.16756,2402.15531,AP463-169617,PDU44-101437,2403.14167,2404.08243,
2405.02328}. However, the topological classes of the multi-charge static AdS black holes in four- and five-dimensional gauged supergravities are still
unknown and merit further exploration, since the structure of the metric, though spherically symmetric, is notably different from
that of the corresponding Reissner-Nordstr\"om-AdS (RN-AdS) cases. Hence the reason why we undertake the present paper.

In this paper, we shall investigate the topological numbers of the static multi-charge AdS black holes in four- and five-dimensional gauged supergravity theories. In the context of gauged supergravity theory, static-charged AdS black holes in four and five dimensions have four and three independent electric charge parameters, respectively. For each of these black holes, we examine various electric charge parameter configurations and explore their impact on the thermodynamic topological classification of these black holes. We find that, in four-dimensional spacetimes, for two nonzero electric charge parameters (the other two being zero), the thermodynamic topological number is temperature dependent: it is $W = 1$ for two large electric charge parameters, but for two small electric charge parameters, it can be $W = 0$ (at cold temperatures) or $W = 1$ (at high temperatures). Likewise, in five-dimensional spacetimes, we also find that the static-charged AdS black hole in Kaluza-Klein (K-K) supergravity theory has $W = 1$ if the electric charge parameter is large, but if the electric charge parameter is small, then the topological number $W$ exhibits a similar temperature dependence, i.e., $W = 0$ (at cold temperatures) or $W = 1$ (at high temperatures). In other words, we observe a kind of novel temperature-dependent thermodynamic topological phase transition.

The remaining part of this paper is organized as follows. In Sec. \ref{II}, we present a brief review of the thermodynamic topological method proposed in Ref. \cite{PRL129-191101}. In Sec. \ref{III}, we examine the topological numbers of four-dimensional static multi-charge AdS black holes in gauged supergravity theory \cite{NPB554-237} for several different combinations of electric charge parameters, and we address each case separately in six subsections. In Sec. \ref{IV}, we investigate the topological numbers of five-dimensional static multi-charge AdS black holes in gauged supergravity theory \cite{NPB553-317} with various distinct combinations of electric charge parameters and separate our discussion of each case into five subsections. Finally, our conclusions and outlooks are given in Sec. \ref{V}.

\section{A brief review of thermodynamic topological method}\label{II}
In this section, we present a brief review of the novel thermodynamic topological method proposed in Ref. \cite{PRL129-191101}. As stated
in Ref. \cite{PRL129-191101}, we start by introducing the generalized off-shell Helmholtz free energy
\be\label{FE}
\mathcal{F} = M -\frac{S}{\tau}
\ee
for the black hole thermodynamic system with mass $M$ and Bekenstein-Hawking entropy $S$, the extra variable $\tau$ can be treated as the inverse
temperature of the cavity enclosing the black hole. Only in the case of $\tau = 1/T$ does the generalized Helmholtz free energy exhibit on-shell
features and return to the standard Helmholtz free energy $F = M -TS$ of the black hole \cite{PRD15-2752,PRD33-2092,PRD105-084030,PRD106-106015}.

In Ref. \cite{PRL129-191101}, the essential vector $\phi$ is defined as
\bea\label{vector}
\phi = \Big(\frac{\p \mathcal{F}}{\p r_h}\, , ~ -\cot\Theta\csc\Theta\Big) \, ,
\eea
where $r_h$ is the event horizon radius of the black hole, $\Theta$ is an extra factor, and $\Theta\in [0,+\infty]$. It is worth noting that the component $\phi^\Theta$ diverges at $\Theta = 0$ and $\Theta = \pi$, demonstrating that the vector has an outward direction in both of these cases.

In order to build a topological current, one can employ Duan's theory \cite{SS9-1072,NPB514-705,PRD61-045004} on $\phi$-mapping topological currents
as follows:
\be\label{jmu}
j^{\mu}=\frac{1}{2\pi}\epsilon^{\mu\nu\rho}\epsilon_{ab}\p_{\nu}n^{a}\p_{\rho}n^{b}\, , \qquad
\mu,\nu,\rho=0,1,2,
\ee
where $\p_{\nu}= \p/\p x^{\nu}$ and $x^{\nu}=(\tau,~r_h,~\Theta)$. The normalized vector is formulated as $n = (n^r, n^\Theta)$ with
$n^r = \phi^{r_h}/||\phi||$ and $n^\Theta = \phi^{\Theta}/||\phi||$. It is simple to verify that this topological current is conserved
\be
\p_{\mu}j^{\mu} = 0 \, .
\ee
Using the three-dimensional Jacobian tensor $\epsilon^{ab}J^{\mu}(\phi/x) = \epsilon^{\mu\nu\rho}\p_{\nu}\phi^a\p_{\rho}\phi^b$, we can describe the topological current as a $\delta$-function of the field configuration \cite{PRD102-064039,NPB514-705,PRD61-045004}
\be
j^{\mu}=\delta^{2}(\phi)J^{\mu}\Big(\frac{\phi}{x}\Big) \, .
\ee
This argument clearly shows that $j^\mu$ is nonzero only at the zero points of $\phi^a(x_i)$, i.e., $\phi^a(x_i) = 0$. Finally, the topological number
at the given parameter region $\Sigma$ can be determined by utilizing the following formula:
\be
W = \int_{\Sigma}j^{0}d^2x = \sum_{i=1}^{N}\beta_{i}\eta_{i} = \sum_{i=1}^{N}w_{i}\, ,
\ee
where the positive Hopf index $\beta_i$ denotes the number of loops made by $\phi^a$ in the vector $\phi$-space as $x^{\mu}$ moves around the zero point
$z_i$, the Brouwer degree $\eta_{i}= \mathrm{sign}(J^{0}({\phi}/{x})_{z_i})=\pm 1$, and $w_i$ is the winding number for the $i$th zero point of $\phi$.
In addition, if two different closed curves, $\Sigma_1$ and $\Sigma_2$, intersect at the same zero point of $\phi$, their winding numbers must be the same; if
there is no zero point of $\phi$ within the enclosed region, then the topological number $W = 0$.

It is worth mentioning that the local winding number $w_{i}$ is a key instrument for determining local thermodynamical stability. Positive $w_{i}$ values indicate thermodynamically stable black holes, while negative values indicate unstable ones. The global topological number $W$ denotes the difference between the numbers of thermodynamically stable and unstable black holes with a black hole solution at a given temperature.

\section{Four-dimensional static multi-charge AdS black holes in gauged supergravity theory}\label{III}
In this section, we will investigate the topological numbers of the four-dimensional static multi-charge AdS black holes in gauged supergravity theory \cite{NPB554-237}. For the general static four-charge AdS black hole in four-dimensional STU gauged supergravity theory, whose metric, Abelian gauge potentials, and scalar fields are \cite{NPB554-237}
\bea\label{4d}
ds_4^2 &=& - \prod_{i=1}^4 H_i^{-1/2}\, f dt^2 + \prod_{i=1}^4 H_i^{1/2}\Big(f^{-1}{dr^2} +r^2d\Omega_{2}^2 \Big) \,,\nn\\
A^i &=& \frac{\sqrt{q_i(q_i +2m)}}{2(r +q_i)}\, dt\, ,\qquad
X_i = H_i^{-1}\, \prod_{j=1}^4 H_j^{1/4}\, ,
\eea
where
\be
f = 1 - \frac{2m}{r} + \frac{r^2}{l^2}\prod_{i=1}^4H_i \, , \qquad H_i = 1 +\frac{q_i}{r} \, ,
\ee
in which $l$ is the AdS radius, $m$ and $q_i$ are the mass and four independent electric charge parameters, respectively.

For the static charged AdS black hole metric described by (\ref{4d}), the most general case is that of a solution possessing four independent electric charge parameters. In addition, according to the classification of black hole solutions in Fig. 1 of Ref. \cite{PRD90-025029}, there are several special truncated supergravity solutions: for example, when the electric charge parameters $q_1 = q_2$ and $q_3 = q_4$, this is known as the static charged AdS black hole solution in gauged $-iX^0X^1$-supergravity theory \cite{NPB444-92} (namely, the static pairwise-equal four-charge AdS black hole case); when $q_1 \ne q_2 \ne 0$, $q_3 = q_4 = 0$, namely the four-dimensional static charged AdS Horowitz-Sen black hole solution \cite{PRD53-808,NPB477-449}; when $q_1 = q_2 \ne 0$ and $q_3 = q_4 = 0$, i.e., the static charged AdS black hole solution in Einstein-Maxwell-dilaton-axion (EMDA) gauged supergravity theory \cite{PRL69-1006}; and when $q_1 \ne 0$ and $q_2 = q_3 = q_4 = 0$, i.e., the static charged AdS black hole solution in K-K gauged supergravity; and when $q_1 = q_2 = q_3 = q_4 \ne 0$, which is the familiar RN-AdS black hole case after the coordinate transformation by $\rho = r +q_i$; and so on.

The thermodynamic quantities are given by \cite{PRD84-024037}
\bea\label{therm}
&&M = m  +\frac{1}{4}\sum_{i=1}^4q_i\, ,\quad Q_i= \frac{1}{2}\sqrt{q_i(q_i +2m)}\, , \quad S = \pi\prod_{i=1}^4 (r_h +q_i)^{1/2}\, , \quad T = \frac{f'(r_h)}{4\pi}\prod_{i=1}^4 H_i^{-1/2}(r_h)\, ,  \nn \\
&&\Phi_i = \frac{\sqrt{q_i(q_i +2m)}}{2(r_h +q_i)}\, , \quad P = \frac{3}{8\pi l^2} \, , \quad V = \frac{\pi r_h^3}{3}\prod_{i=1}^4 H_i(r_h)\sum_{j=1}^4\frac{1}{H_j(r_h)} \, .
\eea

It is easy to verify that these mentioned thermodynamic quantities simultaneously satisfy the first law and the Bekenstein-Smarr mass formula
\bea
dM &=& TdS +\sum_{i=1}^4\Phi_idQ_i +VdP \, , \\
M &=& 2TS +\sum_{i=1}^4\Phi_iQ_i -2VP \, .
\eea

Utilizing the definition of the generalized off-shell Helmholtz free energy (\ref{FE}) and substituting the relation $l^2 = 3/(8\pi P)$ \cite{PRD84-024037,CPL23-1096,CQG26-195011}, one can easily obtain
\bea
\mathcal{F} &=& \frac{r_h}{2} +\frac{1}{4}\sum_{i=1}^4q_i +\frac{4\pi P}{3r_h}\prod_{i=1}^4(r_h +q_i)  -\frac{\pi}{\tau}\prod_{i=1}^4\sqrt{r_h +q_i} \qquad
\eea
for the static four-charge AdS black hole in four-dimensional gauged supergravity. Then the components of the vector $\phi$ can be derived as
\bea
\phi^{r_h} &=& \frac{1}{2} +\frac{4\pi P}{3}\Big[q_2q_3 +(q_2 +q_3)q_4 +q_1(q_2 +q_3 +q_4) -\frac{\prod_{i=1}^4q_i}{r_h^2} +2r_h\sum_{i=1}^4q_i +3r_h^2\Big] \nn \\
&&-\frac{\pi}{6\tau\prod_{i=1}^4\sqrt{r_h +q_i}}\bigg\{ 12r_h^3 +9r_h^2\sum_{i=1}^4q_i +6r_h\Big[q_3q_4 +q_2(q_3 +q_4) +q_1(q_2 +q_3+q_4)\Big] \nn \\ &&+3\Big[q_1q_2q_3 +q_2q_3q_4 +q_1(q_2 +q_3)q_4 \Big] \bigg\} \, , \\
\phi^{\Theta} &=& -\cot\Theta\csc\Theta \, .
\eea
By solving the equation: $\phi^{r_h} = 0$, one can compute the zero point of the vector field $\phi^{r_h}$ as
\be\label{tau4d}
\tau = \frac{3\pi r_h^2\Big[q_1q_2q_3 +q_1q_2q_4 +q_1q_3q_4 +q_2q_3q_4 +2Xr_h +3r_h^2\sum_{i=1}^4q_i +4r_h^3\Big]}{\prod_{i=1}^4\sqrt{r_h +q_i}\Big\{3r_h^2 + 8\pi P\big[-\prod_{i=1}^4q_i +Xr_h^2 +2r_h^3\sum_{i=1}^4q_i +3r_h^4\big]\Big\}} \, ,
\ee
where
\be
X =  q_3q_4 +q_2(q_3 +q_4) +q_1(q_2 +q_3 +q_4) \, . \nn
\ee

Note that Eq. (\ref{tau4d}) consistently reduces to the result obtained in the case of the four-dimensional Schwarzschild-AdS black hole \cite{PRD107-084002} when the four independent electric charge parameters $q_i$ vanish. Due to the requirement of considering four independent electric charge parameters, different values of these electric charge parameters correspond to distinct black hole solutions within various truncated supergravity theories. Therefore, we will explore the topological numbers of static charged AdS black holes in several special supergravity theories, respectively.

\subsection{$q_1 \ne 0$, $q_2 = q_3 = q_4 = 0$ case (K-K gauged supergravity)}\label{IIIA}
\begin{figure}[t]
\centering
\includegraphics[width=0.5\textwidth]{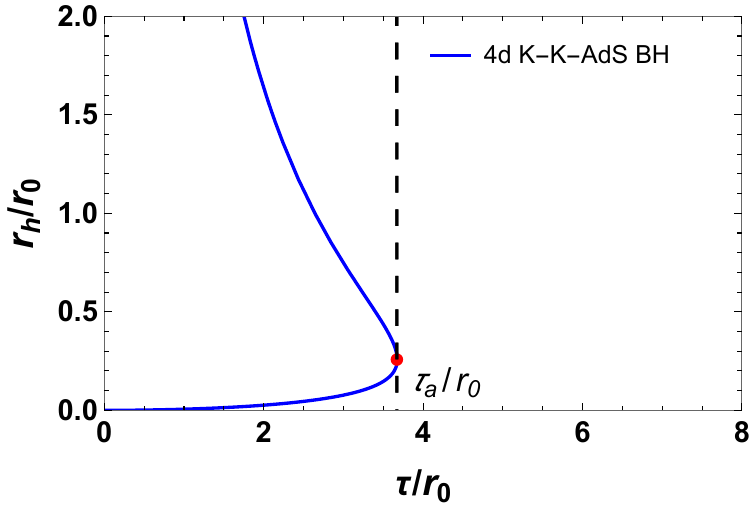}
\caption{Zero points of the vector $\phi^{r_h}$ shown in the $r_h-\tau$ plane with $q_1/r_0 = 2$, $Pr_0^2 = 0.1$, and $q_2 = q_3 = q_4 = 0$. There is one thermodynamically stable and one thermodynamically unstable four-dimensional static charged AdS black hole in K-K gauged supergravity theory for $\tau < \tau_a = 3.68r_0$. Obviously, the topological number is: $W = 1 -1 = 0$.
\label{4d1cBH}}
\end{figure}

\begin{figure}[t]
\centering
\includegraphics[width=0.5\textwidth]{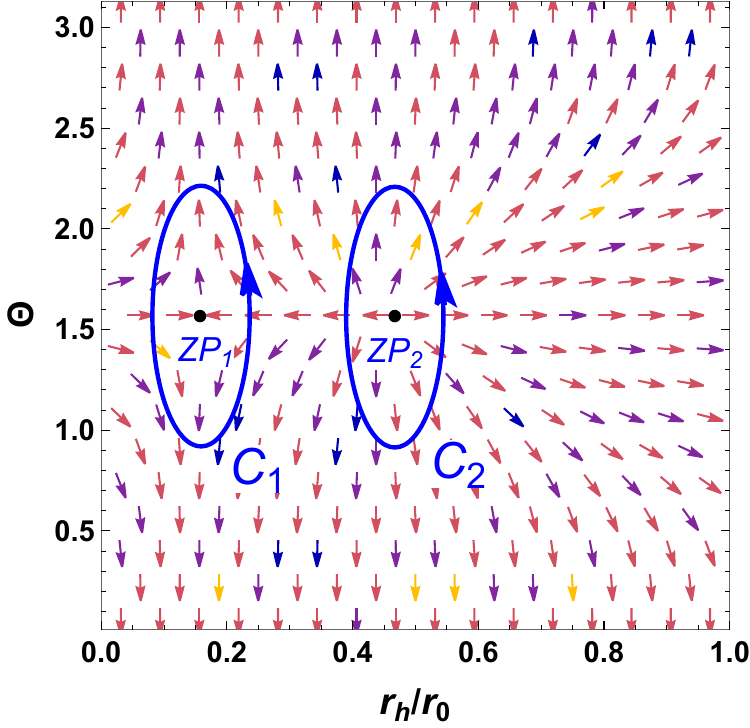}
\caption{The arrows represent the unit vector field $n$ on a portion of the $r_h-\Theta$ plane for the four-dimensional static charged AdS black hole in K-K gauged supergravity theory with $\tau/r_0 = 3.5$, $q_1/r_0 = 2$, $Pr_0^2 = 0.1$, and $q_2 = q_3 = q_4 = 0$. The zero points (ZPs) marked with black dots are at $(r_h/r_0, \Theta) = (0.15,\pi/2)$, and $(0.44,\pi/2)$, respectively. The blue contours $C_i$ are closed loops enclosing the zero points.
\label{d41cBH}}
\end{figure}
In this subsection, we focus on the case where $q_1 \neq 0$ and $q_2 = q_3 = q_4 = 0$, which corresponds to the static charged AdS black hole in four-dimensional K-K gauged supergravity theory. For the four-dimensional static charged AdS black hole in K-K gauged supergravity theory, one can plot the zero points of the component $\phi^{r_h}$ with $Pr_0^2 = 0.1$, $q_1/r_0 = 2$, and $q_2 = q_3 = q_4 = 0$ (the four electric charge parameters act equivalently) in Fig. \ref{4d1cBH}, and the unit vector field $n$ on a portion of the $\Theta-r_h$ plane in Fig. \ref{d41cBH} with $\tau/r_0 = 3.5$. Here, $r_0$ represents an arbitrary length scale defined by the size of a cavity around the static charged AdS black hole in four-dimensional K-K gauged supergravity theory. Figure \ref{4d1cBH} illustrates that for $\tau < \tau_a = 3.68r_0$, there exist two four-dimensional static charged AdS black holes in K-K gauged supergravity: one thermodynamically stable and one thermodynamically unstable.

In Fig. \ref{d41cBH}, two zero points are located at $(r_h/r_0, \Theta) = (0.15,\pi/2)$, and $(0.44,\pi/2)$, respectively. The winding numbers $w_i$ for the blue contours $C_i$ can be characterized as $w_1 = -1$ and $w_2 = 1$, which differ from the four-dimensional RN-AdS black hole (which only has $w_1 = 1$). Therefore, the topological number $W = 0$ for the four-dimensional static charged AdS black hole in K-K gauged supergravity theory is easily noticed in Fig. \ref{d41cBH}, distinguishing it from the topological number of the four-dimensional RN-AdS black hole ($W = 1$) \cite{PRL129-191101}. It implies that the topological number are significantly affected by the number of the electric charge parameters.
\subsection{$q_1 = q_2 \ne 0$, $q_3 = q_4 = 0$ case (EMDA gauged supergravity)}\label{IIIB}

In this subsection, we discuss the case where $q_1 = q_2 \ne 0$, $q_3 = q_4 = 0$, corresponding to the static charged AdS black hole in four-dimensional EMDA gauged supergravity theory. For the four-dimensional static charged AdS black hole in EMDA gauged supergravity theory, we find that different values of the two identical electric charge parameters also influence their topological numbers, which is a new property of this black hole solution (in Secs. \ref{IIIC} and \ref{IVA}, we will find that the four-dimensional static charged AdS Horowitz-Sen black hole solution and the five-dimensional static charged AdS black hole solution in K-K gauged supergravity theory possess similar properties). Therefore, we next discuss each of the three cases by taking larger, smaller, and critical values of two equal electric charge parameters.

\subsubsection{Large values of two identical electric charge parameters}

\begin{figure}[t]
\centering
\includegraphics[width=0.5\textwidth]{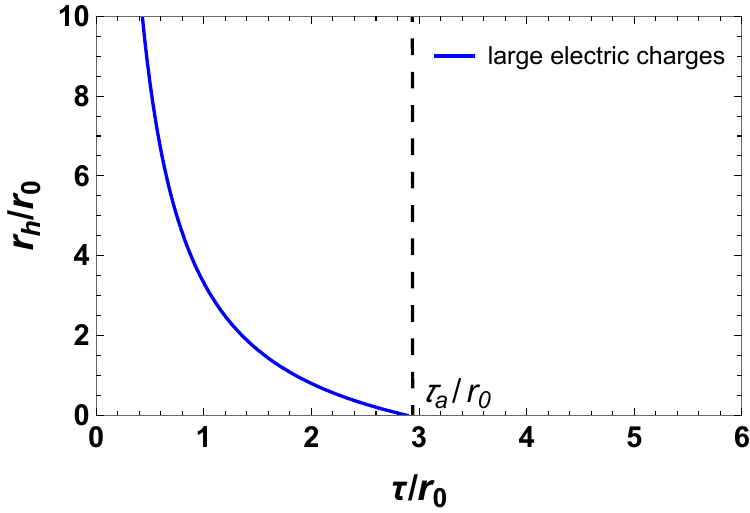}
\caption{Zero points of the vector $\phi^{r_h}$ shown in the $r_h-\tau$ plane with $q_1/r_0 = q_2/r_0 = 2$, $Pr_0^2 = 0.1$, and $q_3 = q_4 = 0$. There is one thermodynamically stable four-dimensional static charged AdS black hole in EMDA gauged supergravity theory for $\tau < \tau_a = 2.89r_0$. Obviously, the topological number is: $W = 1$.
\label{4d2cBHL}}
\end{figure}

\begin{figure}[b]
\centering
\includegraphics[width=0.5\textwidth]{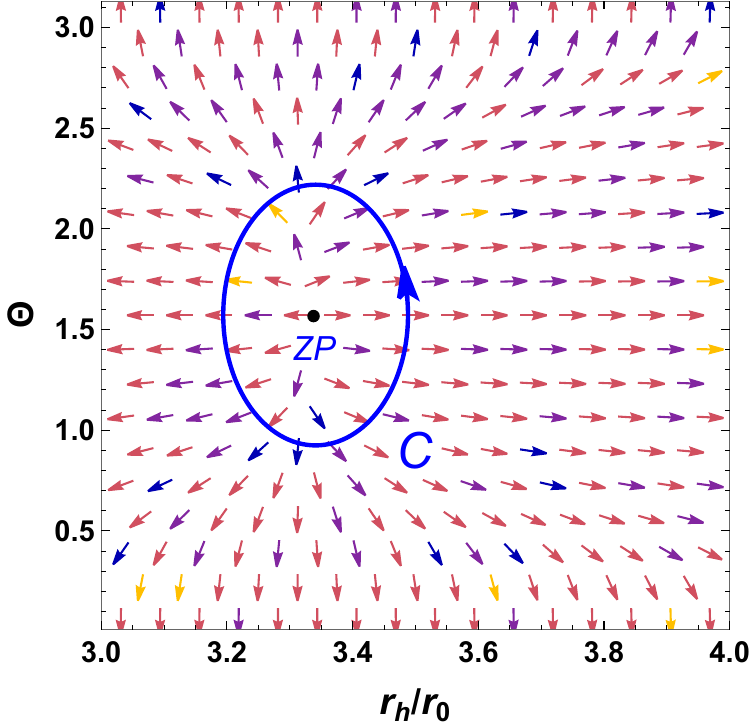}
\caption{The arrows represent the unit vector field $n$ on a portion of the $r_h-\Theta$ plane for the four-dimensional static charged AdS black hole in EMDA gauged supergravity theory with $\tau/r_0 = 1$, $q_1/r_0 = q_2/r_0 = 2$, $Pr_0^2 = 0.1$, and $q_3 = q_4 = 0$. The zero point (ZP) marked with a black dot is at $(r_h/r_0, \Theta) = (3.32,\pi/2)$. The blue contour $C$ is a closed loop enclosing the zero point.
\label{d42cBHL}}
\end{figure}

We first consider the case where two equal electric charge parameters take larger values. We plot the zero points of the component $\phi^{r_h}$ with $Pr_0^2 = 0.1$, $q_1/r_0 = q_2/r_0 = 2$, and $q_3 = q_4 = 0$ in Fig. \ref{4d2cBHL}, and the unit vector field $n$ in Fig. \ref{d42cBHL} with $\tau/r_0 = 1$. Note that for these values of $Pr_0^2$, $q_1/r_0$ and $q_2/r_0$, there is one thermodynamically stable four-dimensional static charged AdS black hole in EMDA gauged supergravity theory if $\tau < \tau_a = 2.89r_0$. In Fig. \ref{d42cBHL}, one can observe that the zero point is located at $(r_h/r_0, \Theta) = (3.32,\pi/2)$. Therefore, the topological number $W = 1$ for the above black hole can be clearly found in Figs. \ref{4d2cBHL} and \ref{d42cBHL} by applying the local property of the zero points, which is the same as that of the four-dimensional RN-AdS black hole \cite{PRL129-191101}, but different from that of the four-dimensional static charged AdS black hole in K-K gauged supergravity theory in the previous subsection, which is $W = 0$.

\subsubsection{Small values of two identical electric charge parameters and the temperature-dependent thermodynamic topological phase transition}\label{IIIB2}

\begin{figure}[t]
\centering
\includegraphics[width=0.5\textwidth]{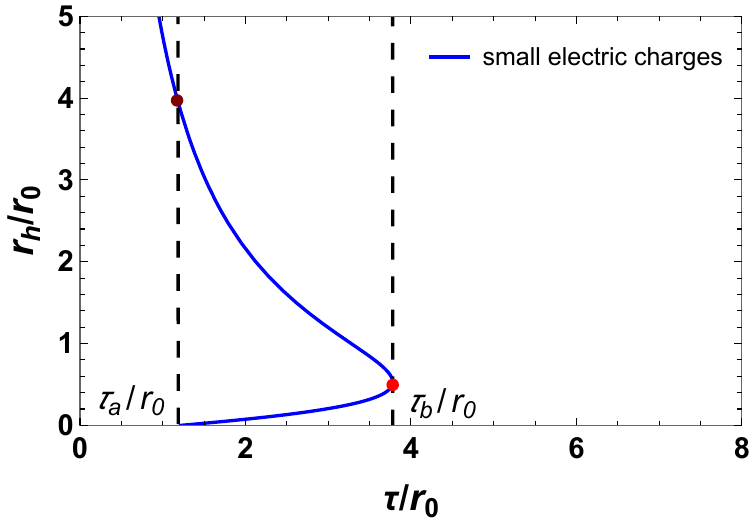}
\caption{Zero points of the vector $\phi^{r_h}$ shown in the $r_h-\tau$ plane with $q_1/r_0 = q_2/r_0 = 0.2$, $q_3 = q_4 = 0$, and $Pr_0^2 = 0.1$. There are one thermodynamically stable and one thermodynamically unstable four-dimensional static charged AdS black hole in EMDA gauged supergravity theory for $1.22r_0 = \tau_a < \tau < \tau_b  = 3.78r_0$, and one thermodynamically stable four-dimensional static charged AdS black hole in EMDA gauged supergravity theory for $\tau < \tau_a = 1.22r_0$.
\label{4d2cBHS}}
\end{figure}

\begin{figure}[t]
\subfigure[~{The unit vector field for the four-dimensional static charged AdS black hole in EMDA gauged supergravity theory with $\tau/r_0 = 3$, $q_1/r_0 = q_2/r_0 = 0.2$, $q_3 = q_4 = 0$, and $Pr_0^2 = 0.1$.}]{\label{d42cBHStau3}
\includegraphics[width=0.5\textwidth]{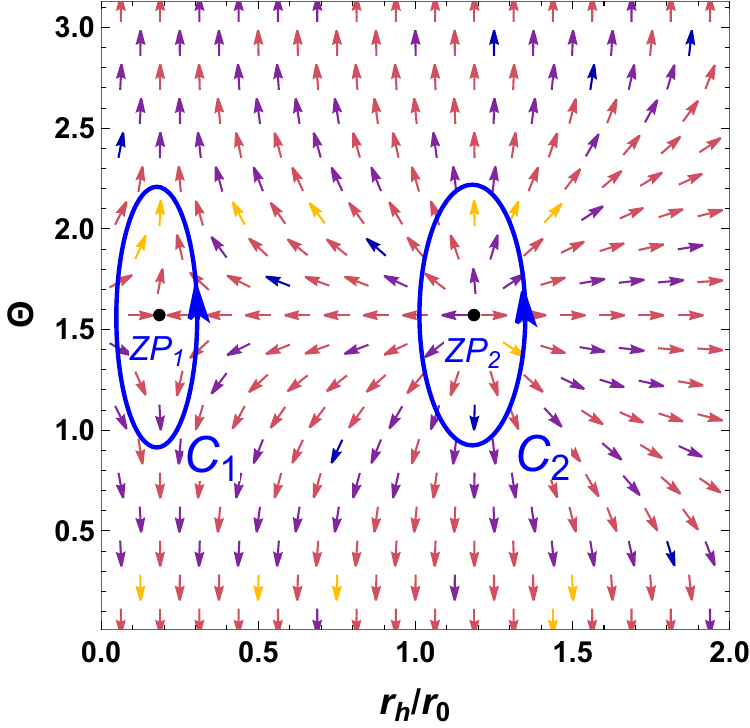}}
\subfigure[~{The unit vector field for the four-dimensional static charged AdS black hole in EMDA gauged supergravity theory with $\tau/r_0 = 1$, $q_1/r_0 = q_2/r_0 = 0.2$, $q_3 = q_4 = 0$, and $Pr_0^2 = 0.1$.}]{\label{d42cBHStau1}
\includegraphics[width=0.5\textwidth]{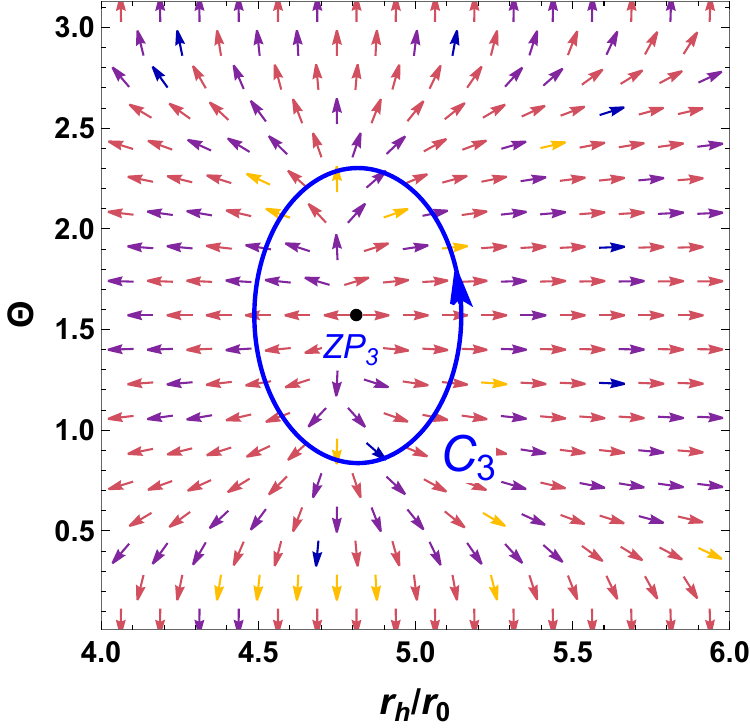}}
\caption{The arrows represent the unit vector field $n$ on a portion of the $r_h-\Theta$ plane. The zero points (ZPs) marked with black dots are at $(r_h/r_0,\Theta) = (0.20,\pi/2)$, $(1.20,\pi/2)$, $(4.75,\pi/2)$, for ZP$_1$, ZP$_2$, and ZP$_3$, respectively. The blue contours $C_i$ are closed loops surrounding the zero points. \label{d42cBHS}}
\end{figure}

Then, we consider the case where two identical electric charge parameters take smaller values. Taking $q_1/r_0 = q_2/r_0 = 0.2$, $q_3 = q_4 = 0$, and $Pr_0^2 = 0.1$, we plot the zero points of the component $\phi^{r_h}$ in Fig. \ref{4d2cBHS}, and the unit vector field $n$ on a portion of the $\Theta-r_h$ plane with $\tau = 3r_0, r_0$ in Fig. \ref{d42cBHS}, respectively. With the help of Fig. \ref{4d2cBHS}, it is easy to figure out that in four-dimensional EMDA gauged supergravity theory, for $1.22r_0 = \tau_a < \tau < \tau_b = 3.78r_0$, there are one thermodynamically stable and one thermodynamically unstable black hole branch, and one thermodynamically stable black hole branch for $\tau < \tau_a = 1.22r_0$. Therefore, the local property of the above black hole for these values of parameters is different from that of the four-dimensional RN-AdS black holes \cite{PRL129-191101}.

Though $(r_h/r_0,\Theta) = (0.20,\pi/2)$ and $(1.20,\pi/2)$, respectively, are the locations of the zero points in Fig.\ref{d42cBHStau3}, for the blue contours $C_i$, one can therefore interpret the winding numbers $w_i$: $w_1 = -1$, $w_2 = 1$, so that the topological number at this inverse temperature $\tau= 3r_0$ is $W = -1 +1 = 0$; However, $(r_h/r_0,\Theta) = (4.75,\pi/2)$ is where the zero point is found in Fig. \ref{d42cBHStau1}, hence the topological number at this inverse temperature $\tau = r_0$ is $W = 1$ since the winding number for the blue contour $C_3$ is $w_3 = 1$. Thus, we find that the topological number is temperature dependent: it is $W = 0$ (at inverse temperature $1.22r_0 = \tau_a < \tau < \tau_b  = 3.78r_0$) or $W = 1$ (at inverse temperature $\tau < \tau_a = 1.22r_0$). At the point of the critical inversion temperature $\tau  = \tau_a$, the black hole occurs a novel temperature-dependent thermodynamic topological phase transition. As can be seen from the Fig. \ref{4d2cBHS}, the function $\tau$ is smooth and continuous at the critical point $\tau_a$, and there is no extreme point nearby $\tau_a$, so the above temperature-dependent thermodynamic topological phase transition is supposed to be a thermodynamic topological higher-order phase transition (continuous phase transition). It is regrettable that we have not yet been able to find an effective method to analyze this higher-order thermodynamic topological phase transition.

\subsubsection{Critical values of two identical electric charge parameters}

The analysis conducted in the preceding two subsubsections indicates that the topological number assumes the value of unity when the magnitudes of the two equivalent electric charge parameters take larger values. Conversely, when these parameters take smaller values, the topological number is temperature-dependent: it is $W = 0$ (at cold temperatures) or $W = 1$ (at high temperatures). It is evident that a critical threshold exists for the two identical electric charge parameters, beyond which the aforementioned transition in the topological number occurs. In other words, there is a topological thermodynamic phase transition at the critical point. In the following, we will investigate the critical value for the two equal electric charge parameters.

\begin{figure}[t]
\centering
\includegraphics[width=0.5\textwidth]{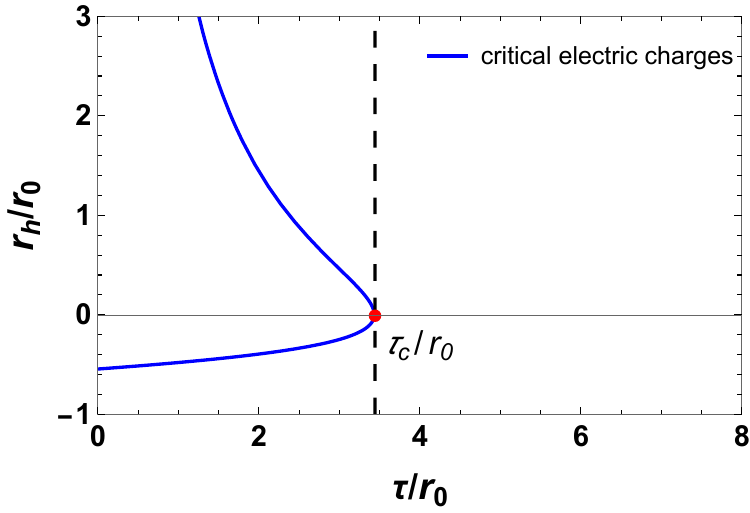}
\caption{Zero points of the vector $\phi^{r_h}$ shown in the $r_h-\tau$ plane with $q_1/r_0 = q_2/r_0 = 1.09$, $q_3 = q_4 = 0$, and $Pr_0^2 = 0.1$. There is one thermodynamically stable four-dimensional static charged AdS black hole in gauged EMDA supergravity theory for $\tau < \tau_c  = 3.43r_0$.
\label{4d2cBHc}}
\end{figure}

\begin{figure}[h]
\centering
\includegraphics[width=0.5\textwidth]{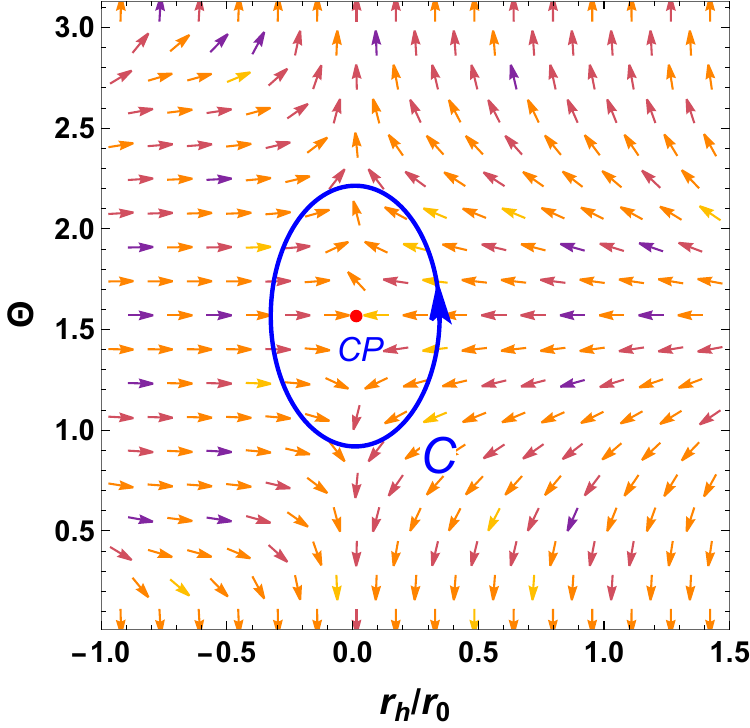}
\caption{The arrows represent the unit vector field $\hat{n}$ on a portion of the $r_h-\Theta$ plane with $q_1/r_0 = q_2/r_0 = 1.09$, $q_3 = q_4 = 0$, and $Pr_0^2 = 0.1$. The critical point (CP) marked with a red dot is at $(r_h/r_0, \Theta) = (0, \pi/2)$. The blue contour $C$ is a closed loop enclosing the critical point.
\label{d42cBHc}}
\end{figure}

When the electric charge parameters $q_1 = q_2 = q$ and $q_3 = q_4 = 0$, the inverse temperature $\tau$ in Eq. (\ref{tau4d}) becomes
\be
\tau = \frac{6\pi(2r_h +q)}{8\pi P(q +3r_h)(q +r_h) +3} \, .
\ee
Now, according to Ref. \cite{PRD105-104003}, one can construct a similar new vector $\varphi = (\varphi^{r_h}, \varphi^\Theta)$
\be\label{varphi}
\varphi^{r_h} = \frac{\p \tau}{\p r_h} \, , \qquad \varphi^{\Theta} = -\cot\Theta\csc\Theta \, .
\ee
The normalized vector field can be obtained through $\hat{n} = (\hat{n}^r, \hat{n}^\Theta)$ with $\hat{n}^r = \varphi^{r_h}/||\varphi||$ and $\hat{n}^\Theta = \varphi^{\Theta}/||\varphi||$. The first advantage of the $\Theta$-term is that the direction of the introduced vector $\varphi$ is vertical to the horizontal lines at $\Theta = 0$ and $\pi$, which can be treated as two boundaries in the parameter space. A further advantage is that the zero point of $\varphi$ is always located at $\Theta = \pi/2$. In addition, it is simple to check that the critical point is located exactly at the zero point of the $\varphi$. Then the components of the vector $\varphi$ can be computed as
\be
\varphi^{r_h} = -\frac{12\pi P[8\pi(3r_h^2 +3qr_h +q^2) -3]}{[8\pi P(q +3r_h)(q +r_h) +3]^2} \, , \qquad \varphi^{\Theta} = -\cot\Theta\csc\Theta \, .
\ee
Therefore, when $r_h \to 0$, the expression for the critical value of the electric charge parameter can be obtained by solving the equation $\varphi^{r_h} = 0$, which is given by
\be\label{qc}
q_c = \frac{\sqrt{6}}{4\sqrt{\pi P}} \, .
\ee
Thus, if $q \ge q_c$, the topological number of the static charged AdS black hole in four-dimensional EMDA gauged supergravity theory is $W = 1$; and when $0 < q < q_c$, the topological number is $W = 0$ (at cold temperatures) or $W = 1$ (at high temperatures).

Taking $q_1/r_0 = q_2/r_0 = q_c/r_0 = 1.09$, $q_3 = q_4 = 0$, and $Pr_0^2 = 0.1$, we plot the zero points of the component $\phi^{r_h}$ in Fig. \ref{4d2cBHc}, and the unit vector field $\hat{n}$ on a portion of the $\Theta-r_h$ plane in Fig. \ref{d42cBHc}, respectively. In Fig. \ref{4d2cBHc}, one can observe that there are one thermodynamically stable four-dimensional static charged AdS black hole in gauged EMDA supergravity theory for $\tau < \tau_c  = 3.43r_0$. In Fig. \ref{d42cBHc}, the critical point (CP) is located at $(r_h/r_0,\Theta) = (0,\pi/2)$, and the topological charge of this critical point is $\hat{W} = -1$, thus it is a conventional critical point \cite{PRD105-104003}.

\subsection{$q_1 \ne q_2 \ne 0$, $q_3 = q_4 = 0$ case ($D = 4$ AdS Horowitz-Sen solution)}\label{IIIC}
In this subsection, we explore a more general solution to the previous subsection, focusing on the case in which the electric charge parameters are $q_1 \ne q_2 \ne 0$ and $q_3 = q_4 = 0$. This specific case corresponds to the four-dimensional static charged AdS Horowitz-Sen black hole solution \cite{1108.5139}. In the following, we first explore whether there is a critical relationship between two different electric charge parameters similar to that of Eq. (\ref{qc}).

\begin{figure}[t]
\centering
\includegraphics[width=0.5\textwidth]{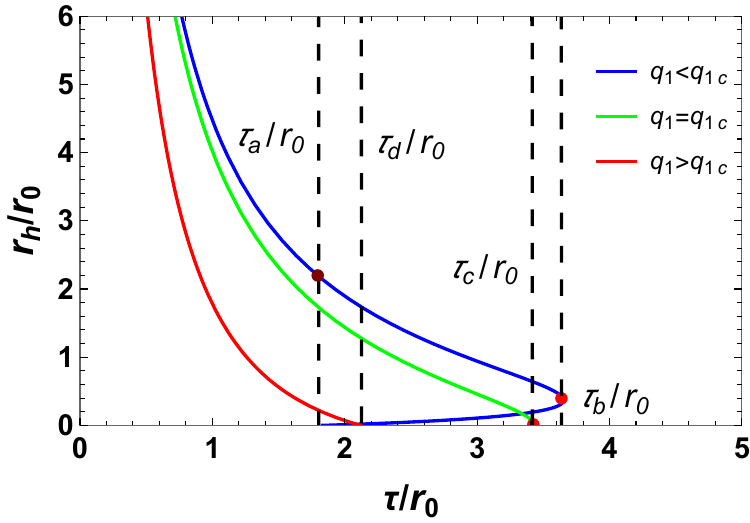}
\caption{Zero points of the vector $\phi^{r_h}$ are depicted in the $r_h-\tau$ plane for the parameters $q_2/r_0 = 1$, $q_3 = q_4 = 0$, and $Pr_0^2 = 0.1$, with three distinct values of $q_1/r_0$: (1) $q_1/r_0 = 0.1$, which is less than the critical value $q_{1c}/r_0$ (represented by the blue solid line); (2) $q_1/r_0 = 15/(4\pi)$, equal to the critical value $q_{1c}/r_0$ (depicted by the green solid line); and (3) $q_1/r_0 = 10$, which exceeds the critical value $q_{1c}/r_0$ (illustrated by the red solid line).
\label{4dHS}}
\end{figure}

When the electric charge parameters $q_1 \ne q_2 \ne 0$ and $q_3 = q_4 = 0$, the inverse temperature $\tau$ in Eq. (\ref{tau4d}) becomes
\be
\tau = \frac{3\pi[4r_h^2 +3r_h(q_1 +q_2) +2q_1q_2]}{\sqrt{r_h +q_1}\sqrt{r_h +q_2}\{8\pi P[3r_h^2 +2r_h(q_1 +q_2) +q_1q_2] +3\}} \, .
\ee
According to the definition of vector $\varphi$ in Eq. (\ref{varphi}), solving the equation $\varphi^{r_h} = 0$ and taking the limit $r_h \to 0$, one can obtain the following critical relationship as
\be\label{q1c}
q_{1c} = \frac{3}{8\pi Pq_2} \, .
\ee
Therefore, for a fixed electric charge parameter $q_2$ and a fixed pressure $P$, if the electric charge parameter $q_1 \ge q_{1c}$, the topological number of the four-dimensional static charged AdS Horowitz-Sen black hole is $W = 1$; and when $0 < q_1 < q_{1c}$, the topological number is $W = 0$ (at cold temperatures) or $W = 1$ (at high temperatures), which is exhibited in Fig. \ref{4dHS} with a fixed electric charge parameter $q_2/r_0 = 1$ and a fixed pressure $Pr_0^2 = 0.1$.

In Fig. \ref{4dHS}, one can observe that when $q_1/r_0 = 10 > q_{1c}/r_0$ (the red solid line), there is one thermodynamically stable four-dimensional static charged Horowitz-Sen AdS black hole for $\tau < \tau_d = 2.12r_0$, and the topological number $W = 1$; when $q_1/r_0 = 15/(4\pi) = q_{1c}/r_0$ (the green solid line), there is one thermodynamically stable four-dimensional static charged Horowitz-Sen AdS black hole for $\tau < \tau_c = 3.41r_0$, and the topological number $W = 1$; when $q_1/r_0 = 0.1 < q_{1c}/r_0$ (the blue solid line),  there are one thermodynamically stable and one thermodynamically unstable four-dimensional static charged Horowitz-Sen AdS black hole branch for $1.80r_0 = \tau_a < \tau < \tau_b = 3.64r_0$, and one thermodynamically stable four-dimensional static charged Horowitz-Sen AdS black hole branch for $\tau < \tau_a = 1.80r_0$, thus the topological number is $W = 0$ (at inverse temperatures $1.80r_0 = \tau_a < \tau < \tau_b = 3.64r_0$) or $W = 1$ (at inverse temperatures $\tau < \tau_a = 1.80r_0$). According to the analysis in the previous subsection, the critical point corresponding to the inverse temperature $\tau_a$ should be a thermodynamic topological higher-order phase transition critical point, while the critical point corresponding to the inverse temperature $\tau_c$ should be a thermodynamic topological conventional phase transition critical point.

At the end of this subsection, we address an important issue. As the smaller electric charge parameter tends to zero, the four-dimensional static charged Horowitz-Sen AdS black hole asymptotically transitions into the four-dimensional static charged AdS black hole in K-K gauged supergravity theory in Sec. \ref{IIIA}. For the latter, the topological number $W$ consistently assumes a single value, $W = 0$. The question arises: is the vanishing of the smaller electric charge parameter a prerequisite for the emergence of a single value for the topological number, or is there a critical threshold for the smaller electric charge parameter below which a temperature-dependent thermodynamic topological phase transition does not happen? To ensure that the topological number for the four-dimensional static charged AdS Horowitz-Sen black hole assumes a single value, it is imperative to satisfy the specific constraint stated in Eq. (\ref{q1c}), namely,
\be
q_1 \ge q_{1c} = \frac{3}{8\pi Pq_2} \, .
\ee
It is readily apparent that when $q_2 = 0$, the value of $q_1$ becomes infinite. Consequently, it is essential for one of the electric charge parameters to be zero in order to obtain a single value for the topological number.

\subsection{$q_1 = q_2 = q_3 = q_4 \ne 0$ case (RN-AdS$_4$)}

\begin{figure}[b]
\centering
\includegraphics[width=0.5\textwidth]{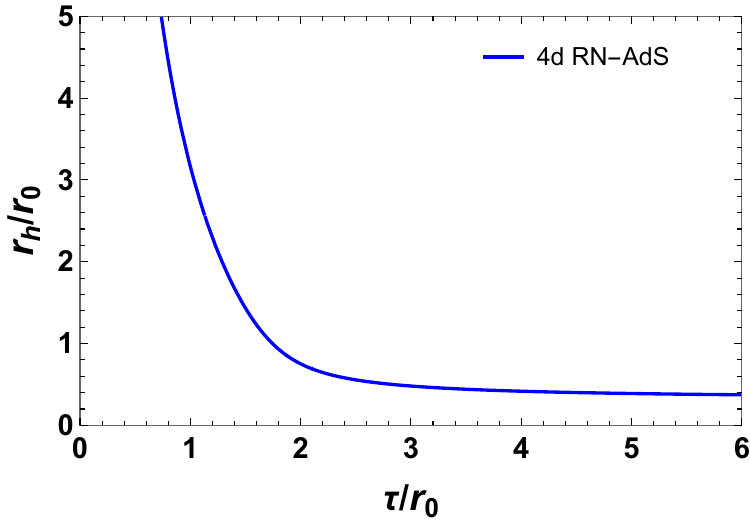}
\caption{Zero points of the vector $\phi^{r_h}$ shown in the $r_h-\tau$ plane with $q_1/r_0 = q_2/r_0 = q_3/r_0 = q_4/r_0 = 1$, and $Pr_0^2 = 0.1$. There is one thermodynamically stable four-dimensional RN-AdS black hole for any value of $\tau$.
\label{4dRNAdS}}
\end{figure}

\begin{figure}[t]
\centering
\includegraphics[width=0.5\textwidth]{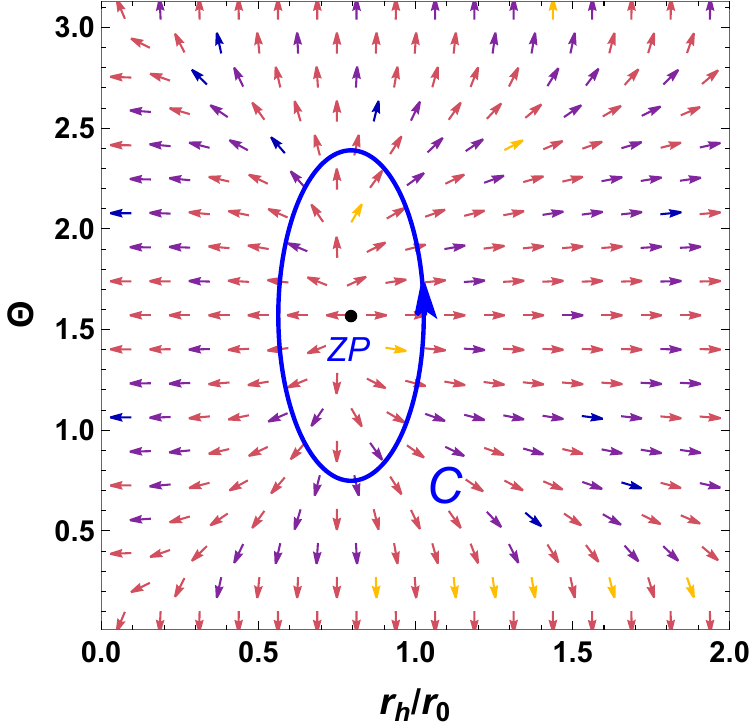}
\caption{The arrows represent the unit vector field $n$ on a portion of the $r_h-\Theta$ plane for the four-dimensional RN-AdS black hole with $\tau/r_0 = 2$, $q_1/r_0 = q_2/r_0 = q_3/r_0 = q_4/r_0 = 1$, $Pr_0^2 = 0.1$. The zero point (ZP) marked with a black dot is at $(r_h/r_0, \Theta) = (0.75,\pi/2)$. The blue contour $C$ is a closed loop enclosing the zero point.
\label{d4RNAdS}}
\end{figure}

Considering the pressure as $Pr_0^2 = 0.1$ and the four electric charge parameters $q_1/r_0 = q_2/r_0 = q_3/r_0 = q_4/r_0 = 1$ for the four-dimensional RN-AdS black hole, we plot the zero points of $\phi^{r_h}$ in the $r_h-\tau$ plane in Fig. \ref{4dRNAdS}, and the unit vector field $n$ on a portion of the $\Theta-r_h$ plane with $\tau/r_0 = 2$ in Fig. \ref{d4RNAdS}, respectively. Based upon the local property of the zero point, one can easily find that the topological number is: $W = 1$, which is consistent with the result given in Ref. \cite{PRL129-191101}.

\subsection{$q_1 = q_2 \ne 0$, $q_3 = q_4 \ne 0$ case (pairwise-equal AdS)}

\begin{figure}[h]
\centering
\includegraphics[width=0.5\textwidth]{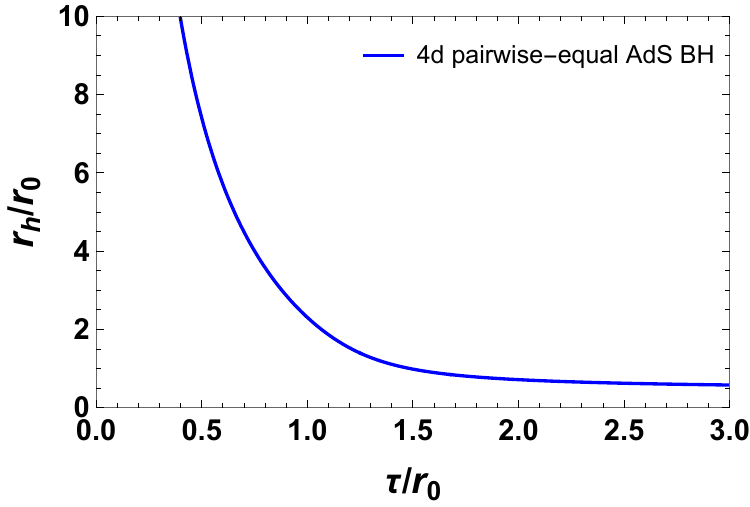}
\caption{Zero points of the vector $\phi^{r_h}$ shown in the $r_h-\tau$ plane with $q_1/r_0 = q_2/r_0 = 2$, $q_3/r_0 = q_4/r_0 = 1$, and $Pr_0^2 = 0.1$. There is always one thermodynamically stable four-dimensional static pairwise-equal four-charge AdS black hole in gauged $-iX^0X^1$-supergravity theory for any value of $\tau$. Obviously, the topological number is: $W = 1$.
\label{4d4cBH}}
\end{figure}

\begin{figure}[h]
\centering
\includegraphics[width=0.5\textwidth]{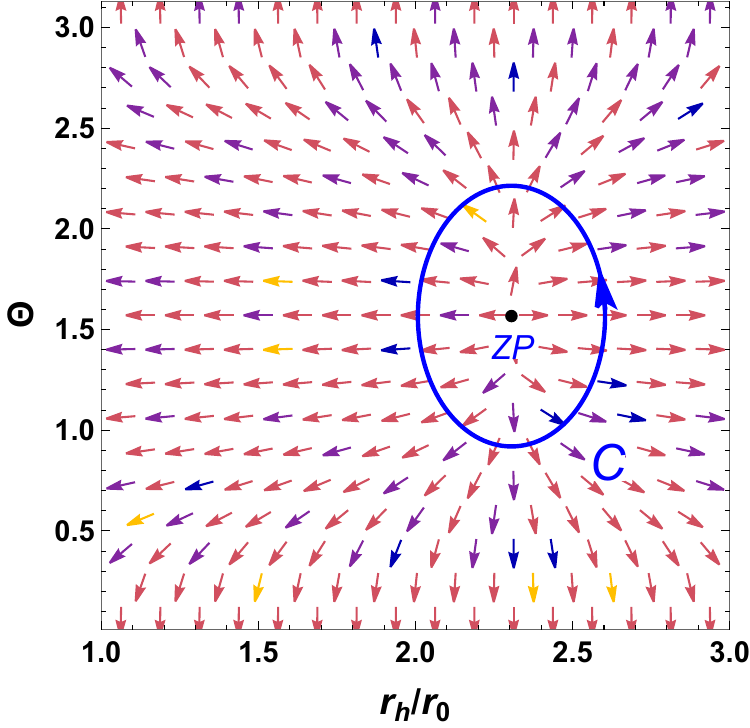}
\caption{The arrows represent the unit vector field $n$ on a portion of the $r_h-\Theta$ plane for the four-dimensional static pairwise-equal four-charge AdS black hole in gauged $-iX^0X^1$-supergravity theory with $\tau/r_0 = 1$, $q_1/r_0 = q_2/r_0 = 2$, $q_3/r_0 = q_4/r_0 = 1$, and $Pr_0^2 = 0.1$. The zero point (ZP) marked with a black dot is at $(r_h/r_0, \Theta) = (2.31,\pi/2)$. The blue contour $C$ is a closed loop enclosing the zero point.
\label{d44cBH}}
\end{figure}

In this subsection, we investigate a special case of four electric charge parameters: $q_1 = q_2 \ne 0$ and $q_3 = q_4 \ne 0$, characterizing the four-dimensional static pairwise-equal four-charge AdS black hole in gauged $-iX^0X^1$-supergravity theory. In Figs. \ref{4d4cBH} and \ref{d44cBH}, taking $q_1/r_0 = q_2/r_0 = 2$, $q_3/r_0 = q_4/r_0 = 1$, and $Pr_0^2 = 0.1$ for the four-dimensional static pairwise-equal four-charge AdS black hole in gauged $-iX^0X^1$-supergravity theory, we plot the zero points of $\phi^{r_h}$ in the $r_h-\tau$ plane and the unit vector field $n$ with $\tau = r_0$, respectively. In Fig. \ref{4d4cBH}, one can observe that there is always one thermodynamically stable four-dimensional static pairwise-equal four-charge AdS black hole in gauged $-iX^0X^1$-supergravity theory for any value of $\tau$. In Fig. \ref{d44cBH}, We have one zero point at $(r_h/r_0, \Theta) = (2.31,\pi/2)$. Based on the local property of the zero points, it is easy to find that the topological number $W = 1$ for the four-dimensional static pairwise-equal four-charge AdS black hole in gauged $-iX^0X^1$-supergravity theory.

\subsection{$q_1 \ne q_2 \ne q_3 \ne q_4 \ne 0$ case (STU gauged supergravity)}\label{3F}

\begin{figure}[b]
\centering
\includegraphics[width=0.5\textwidth]{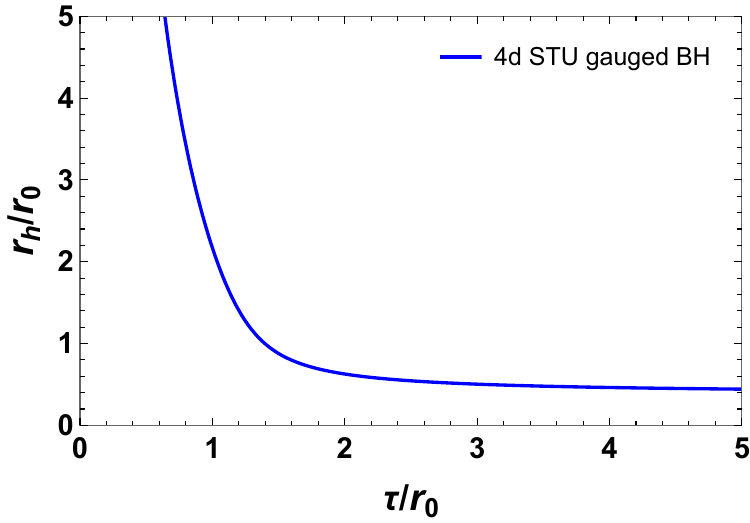}
\caption{Zero points of the vector $\phi^{r_h}$ shown in the $r_h-\tau$ plane with $q_1/r_0 = 0.5$, $q_2/r_0 = 1$, $q_3/r_0 = 2$, $q_4/r_0 = 3$, and $Pr_0^2 = 0.1$. There is always one thermodynamically stable four-dimensional static four-charge AdS black hole in STU gauged supergravity theory for any value of $\tau$. Obviously, the topological number is: $W = 1$.
\label{4dSTU}}
\end{figure}

\begin{figure}[t]
\centering
\includegraphics[width=0.5\textwidth]{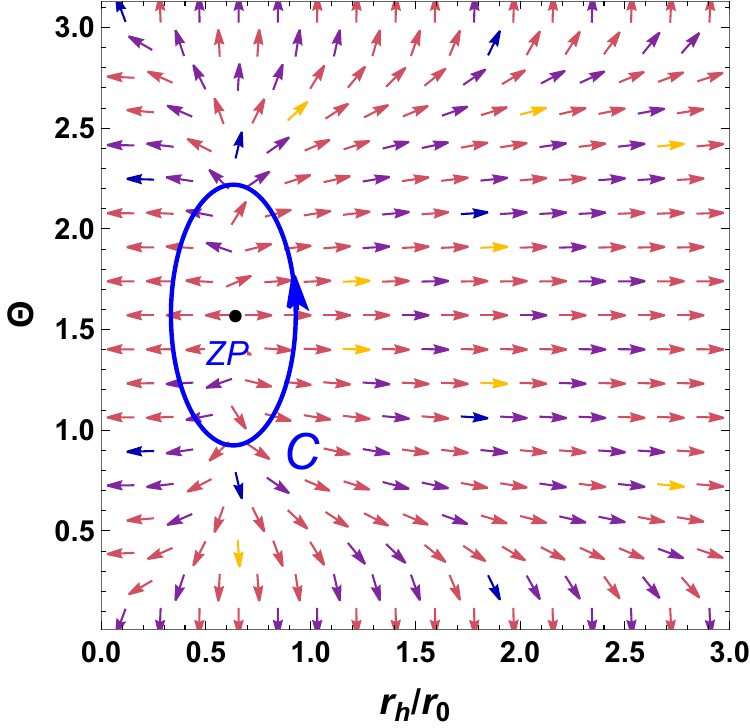}
\caption{The arrows represent the unit vector field $n$ on a portion of the $r_h-\Theta$ plane for the four-dimensional static four-charge AdS black hole in STU gauged supergravity theory with $\tau/r_0 = 2$, $q_1/r_0 = 0.5$, $q_2/r_0 = 1$, $q_3/r_0 = 2$, $q_4/r_0 = 3$, and $Pr_0^2 = 0.1$. The zero point (ZP) marked with a black dot is at $(r_h/r_0, \Theta) = (0.63,\pi/2)$. The blue contour $C$ is a closed loop enclosing the zero point.
\label{d4STU}}
\end{figure}

In this subsection, we consider the most general static four-charge AdS black hole case in STU gauged supergravity theory, i.e., $q_1 \ne q_2 \ne q_3 \ne q_4 \ne 0$ case. We take $q_1/r_0 = 0.5$, $q_2/r_0 = 1$, $q_3/r_0 = 2$, $q_4/r_0 = 3$, and $Pr_0^2 = 0.1$, and then plot the zero points of the component $\phi^{r_h}$ in Fig. \ref{4dSTU}, and the unit vector field $n$ on a portion of the $\Theta-r_h$ plane with $\tau/r_0 = 2$ in Fig. \ref{d4STU}, respectively. It is easy to observe that there is always one thermodynamically stable four-dimensional static four-charge AdS black hole in STU gauged supergravity theory for any value of $\tau$. In Fig. \ref{d4STU}, we observe a zero point at $(r_h/r_0, \Theta) = (0.63,\pi/2)$. Based upon the local property of the zero points, it is simple to demonstrate that the topological number $W = 1$ for the four-dimensional static four-charge AdS black hole in STU gauged supergravity theory.

\section{Five-dimensional static multi-charge black holes in gauged supergravity theory}\label{IV}
In this section, we would like to investigate the topological numbers of the five-dimensional static multi-charge AdS black holes in gauged supergravity theory \cite{NPB553-317}.  For the general static three-charge AdS black hole in five-dimensional STU gauged supergravity theory, whose metric, Abelian gauge potentials, and scalar fields are \cite{NPB553-317}
\bea\label{5d}
ds_5^2 &=& -\prod_{i=1}^3 H_i^{-2/3}fdt^2 +\prod_{i=1}^3 H_i^{1/3}\bigg(f^{-1}{dr^2}
+r^2d\Omega_{3}^2 \bigg) \, , \nn \\
A^i &=& \frac{\sqrt{q_i(q_i +2m)}}{r^2 +q_i}\, dt\, ,\qquad
X_i = H_i^{-1}\prod_{j=1}^3 H_j^{1/3}\, ,
\eea
where
\be
f = 1 -\frac{2m}{r^2} +\frac{r^2}{l^2}\prod_{i=1}^3H_i \, ,
\qquad H_i = 1 +\frac{q_i}{r^2} \, ,
\ee
in which $l$ is the AdS radius, $m$ and $q_i$ are the mass and three independent electric charge parameters, respectively.

For the metric of a five-dimensional static, charged AdS black hole, as expressed in Eq. (\ref{5d}), the most general case is represented by a solution with three independent electric charge parameters. Moreover, in line with the classification scheme for black hole solutions depicted in Fig. 1 of Ref. \cite{PRD90-025029}, many specific truncated supergravity solutions are identified: for instance, when $q_1 \ne 0$ and $q_2 = q_3 = 0$, namely, the five-dimensional static charged AdS black hole solution in K-K gauged supergravity; when $q_1 = q_2 \ne 0$ and $q_3 = 0$, i.e., the five-dimensional static charged AdS black hole solution in EMDA gauged supergravity theory; when $q_1 \ne q_2 \ne 0$ and $q_3 = 0$, i.e., the five-dimensional static charged AdS Horowitz-Sen black hole solution \cite{1108.5139}; and when $q_1 = q_2 = q_3 \ne 0$, which is the famous five-dimensional RN-AdS black hole case after the coordinate transformation by $\rho^2 = r^2 + q_i$; etc.

The thermodynamic quantities are \cite{PRD84-024037}
\be\ba\label{Therm}
&M = \frac{3\pi}{4}m +\frac{\pi}{4}\sum_{i=1}^3q_i  \, ,\quad Q_i= \frac{1}{4}\pi \sqrt{q_i(q_i +2m)} \, , \quad S = \frac{1}{2}\pi^2\prod_{i=1}^3(r_h^2 +q_i)^{1/2}\, , \quad P = \frac{3}{4\pi l^2} \, , \\
&T = \frac{f'(r_h)}{4\pi}\prod_{i=1}^3 H_i^{-1/2}(r_h)\, , \quad \Phi_i = \frac{\sqrt{q_i(q_i +2m)}}{r_h^2 +q_i}\, , \quad V = \frac{\pi^2r_h^4}{6}\prod_{i=1}^3 H_i(r_h)\sum_{j=1}^3 \frac{1}{H_j(r_h)}\, .
\ea\ee
Then one can verify that the above thermodynamic quantities completely obey the first law and the Bekenstein-Smarr mass formula simultaneously,
\bea
dM &=& TdS +\sum_{i=1}^3\Phi_idQ_i +VdP \, , \\
2M &=& 3TS +2\sum_{i=1}^3\Phi_iQ_i -2VP \, .
\eea

From Eq. (\ref{Therm}), one can obtain the expression of the generalized Helmholtz free energy as
\bea
\mathcal{F} &=& \frac{\pi}{4}\left[\frac{2\pi P\prod_{i=1}^3(r_h^2 +q_i)}{r_h^2} +\frac{3}{2}r_h^2 +\sum_{i=1}^3q_i \right] \nn \\
&&-\frac{\pi^2\prod_{i=1}^3\sqrt{r_h^2 +q_i}}{2\tau} \, . \qquad
\eea
Using the definition of Eq. (\ref{vector}), the components vector $\phi$ can be easily obtained as follows:
\bea
\phi^{r_h} &=& -\frac{3\pi^2r_h^5 +2\pi^2r_h^3\sum_{i=1}^3q_i +\pi^2r_h\big[q_2q_3 +q_1(q_2 +q_3)\big]}{2\tau\prod_{i=1}^3\sqrt{r_h^2 +q_i}} \nn \\
&&+\frac{\pi^2P\big(2r_h^6 +r_h^4\sum_{i=1}^3q_i -\prod_{i=1}^3q_i\big)}{r_h^3}
+\frac{3\pi r_h}{4} \, , \\
\phi^{\Theta} &=& -\cot\Theta\csc\Theta \, .
\eea
It is simple to obtain
\be\label{tau5d}
\tau = \frac{2\pi r_h^4\Big(3r_h^4 +2r_h^2\sum_{i=1}^3q_i +q_1q_2 +q_1q_3 +q_2q_3 \Big)}{3r_h^4 +4\pi P\Big(2r_h^6 +r_h^4\sum_{i=1}^3q_i -\prod_{i=1}^3q_i \Big)\prod_{i=1}^3\sqrt{r_h^2 +q_i}}
\ee
as the zero point of the vector field $\phi$, which consistently reduces to the one obtained in the five-dimensional Schwarzschild-AdS black hole case
when the three independent electric charge parameters are turned off.

Similar to Sec. \ref{III}, varying the three independent electric charge parameters yields distinct black hole solutions within various truncated supergravity theories. In the following, we will investigate the topological numbers of static, charged AdS black holes in some famous five-dimensional supergravity theories, respectively.

\subsection{$q_1 \ne 0$, $q_2 = q_3 = 0$ case (K-K gauged supergravity)}\label{IVA}
In this subsection, we focus on the case where $q_1 \ne 0$ and $q_2 = q_3 = 0$, which corresponds to the static charged AdS black hole in five-dimensional K-K gauged supergravity theory. For the five-dimensional static charged AdS black hole in K-K gauged supergravity theory, similar to the cases of the four-dimensional static charged AdS black hole in EMDA gauged supergravity theory in Sec. \ref{IIIB} and the four-dimensional static charged AdS Horowitz-Sen black hole in Sec. \ref{IIIC}, we find that different values of the electric charge parameter also influence its topological number. Therefore, we also discuss each of the three cases by taking the larger, smaller, and critical values of the electric charge parameter.

\subsubsection{Large value of electric charge parameter}\label{IVA1}

\begin{figure}[t]
\centering
\includegraphics[width=0.5\textwidth]{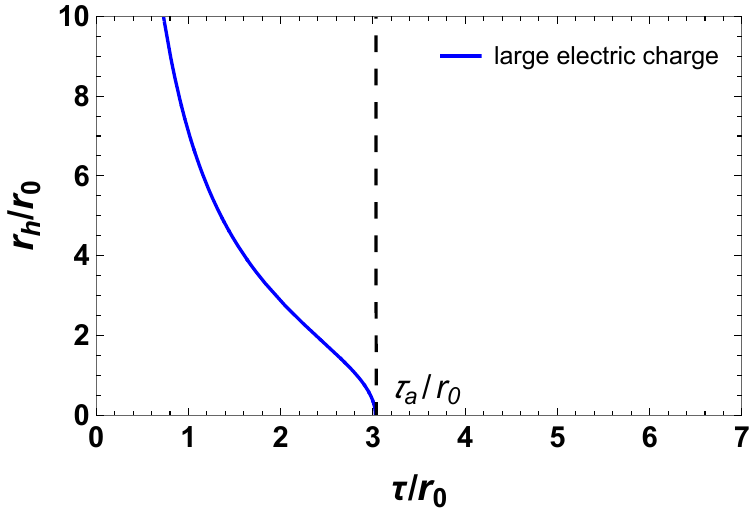}
\caption{Zero points of the vector $\phi^{r_h}$ shown in the $r_h-\tau$ plane with $q_1/r_0^2 = 5$, $Pr_0^2 = 0.1$, and $q_2 = q_3 = 0$. There is one thermodynamically stable five-dimensional static charged AdS black hole in K-K gauged supergravity theory for $\tau < \tau_a = 3.03r_0$. Obviously, the topological number is: $W = 1$.
\label{5d1cBHL}}
\end{figure}

\begin{figure}[h]
\centering
\includegraphics[width=0.5\textwidth]{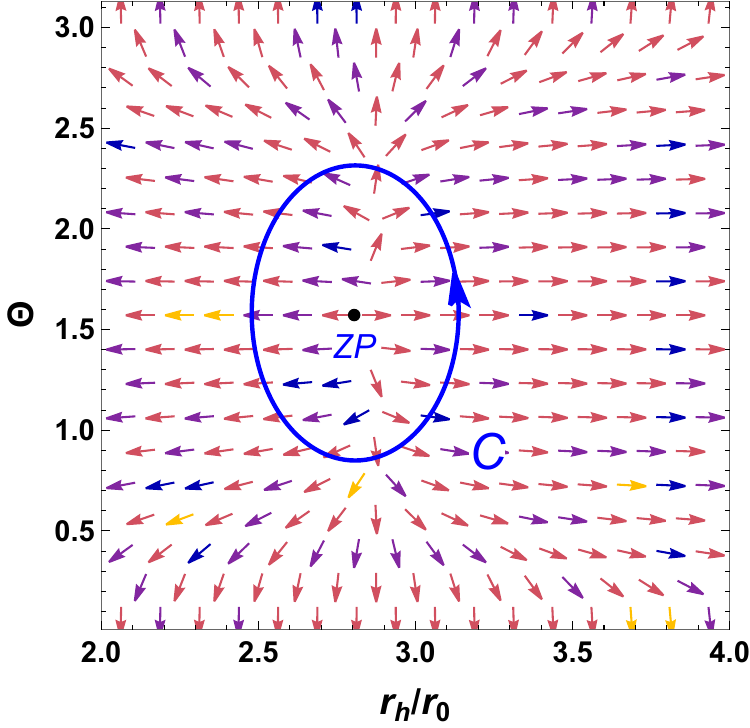}
\caption{The arrows represent the unit vector field $n$ on a portion of the $r_h-\Theta$ plane for the five-dimensional static charged AdS black hole in K-K gauged supergravity theory with $\tau/r_0 = 2$, $q_1/r_0^2 = 5$, $Pr_0^2 = 0.1$, and $q_2 = q_3 = 0$. The zero point (ZP) marked with a black dot is at $(r_h/r_0, \Theta) = (2.87,\pi/2)$. The blue contour $C$ is a closed loop enclosing the zero point.
\label{d51cBHL}}
\end{figure}

We first consider the case where the electric charge parameter takes a larger value. We plot the zero points of the component $\phi^{r_h}$ with $Pr_0^2 = 0.1$, $q_1/r_0^2 = 5$, and $q_2 = q_3 = 0$ in Fig. \ref{5d1cBHL}, and the unit vector field $n$ in Fig. \ref{d51cBHL} with $\tau/r_0 = 2$. Note that for these values of $Pr_0^2$ and $q_1/r_0^2$, there is one thermodynamically stable five-dimensional static charged AdS black hole in K-K gauged supergravity theory for $\tau < \tau_a = 3.03r_0$. In Fig. \ref{d51cBHL}, one can observe that the zero point is located at $(r_h/r_0, \Theta) = (2.87,\pi/2)$. Therefore, the topological number $W = 1$ for the above black hole can be clearly found in Figs. \ref{5d1cBHL} and \ref{d51cBHL} by applying the local property of the zero point, which is the same as that of the five-dimensional RN-AdS black hole \cite{PRL129-191101}.

\subsubsection{Small value of electric charge parameter and the temperature-dependent thermodynamic topological phase transition}\label{IVA2}

\begin{figure}[t]
\centering
\includegraphics[width=0.5\textwidth]{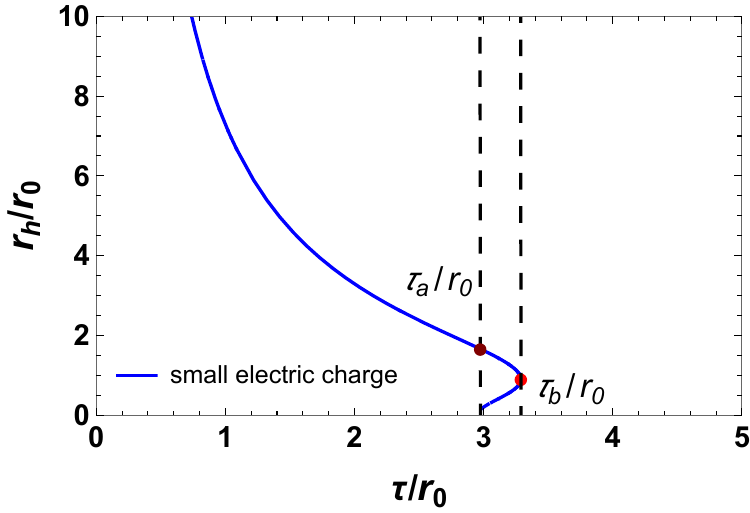}
\caption{Zero points of the vector $\phi^{r_h}$ shown in the $r_h-\tau$ plane with $q_1/r_0^2 = 1$, $q_2 = q_3 = 0$, and $Pr_0^2 = 0.1$. There is one thermodynamically stable and one thermodynamically unstable five-dimensional static charged AdS black hole in K-K gauged supergravity theory for $2.95r_0 = \tau_a < \tau < \tau_b  = 3.29r_0$, and one thermodynamically stable five-dimensional static charged AdS black hole in K-K gauged supergravity theory for $\tau < \tau_a = 2.95r_0$.
\label{5d1cBH}}
\end{figure}

\begin{figure}[t]
\subfigure[~{The unit vector field for the five-dimensional static charged AdS black hole in K-K gauged supergravity theory with $\tau/r_0 = 3.2$, $q_1/r_0^2 = 1$, $q_2 = q_3 = 0$, and $Pr_0^2 = 0.1$.}]{\label{d51cBHtau32}
\includegraphics[width=0.5\textwidth]{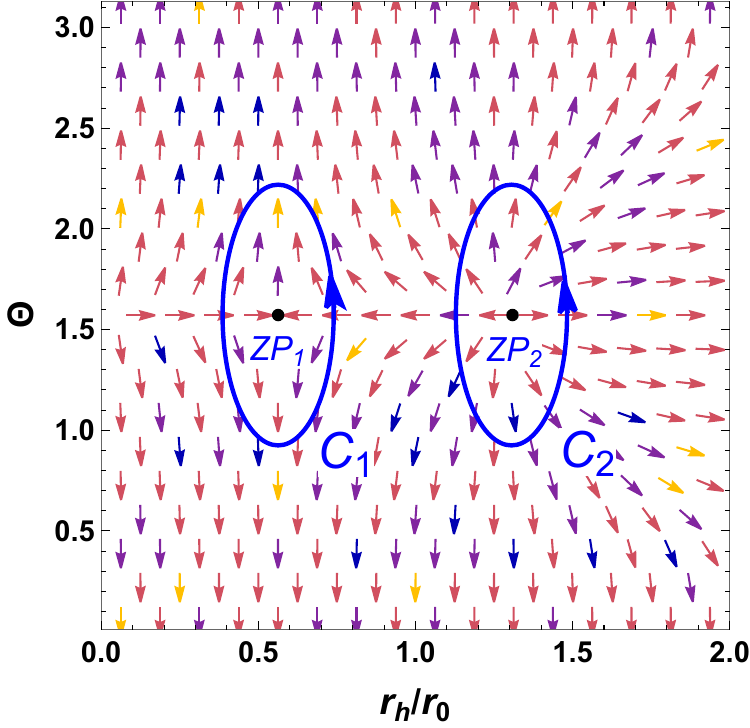}}
\subfigure[~{The unit vector field for the five-dimensional static charged AdS black hole in K-K gauged supergravity theory with $\tau/r_0 = 2$, $q_1/r_0^2 = 1$, $q_2 = q_3 = 0$, and $Pr_0^2 = 0.1$.}]{\label{d51cBHtau2}
\includegraphics[width=0.5\textwidth]{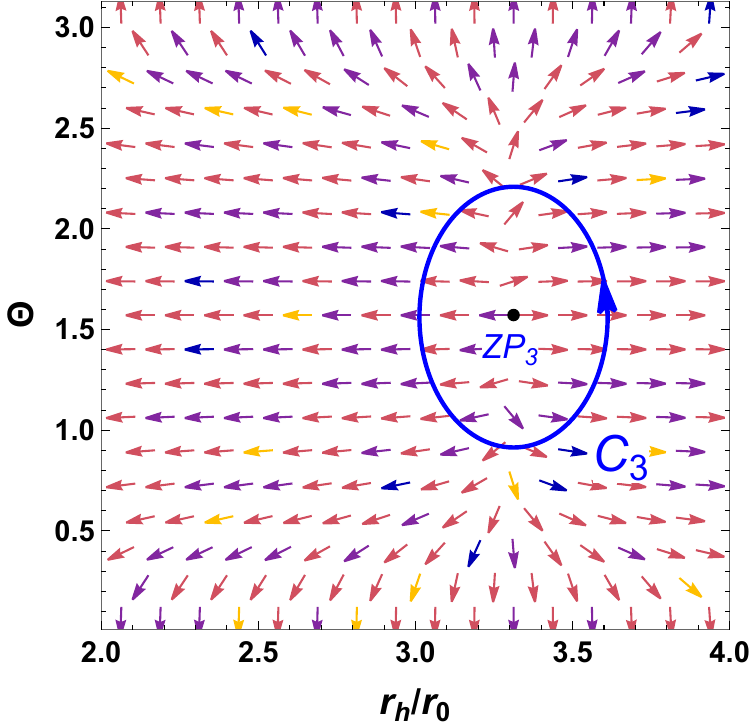}}
\caption{The arrows represent the unit vector field $n$ on a portion of the $r_h-\Theta$ plane. The zero points (ZPs) marked with black dots are at $(r_h/r_0,\Theta) = (0.57,\pi/2)$, $(1.27,\pi/2)$, $(3.29,\pi/2)$, for ZP$_1$, ZP$_2$, and ZP$_3$, respectively. The blue contours $C_i$ are closed loops surrounding the zero points. \label{d51cBH}}
\end{figure}

Then, we consider the case where the electric charge parameter takes a smaller value. We take $q_1/r_0^2 = 1$, $q_2 = q_3 = 0$, and $Pr_0^2 = 0.1$, and then plot the zero points of the component $\phi^{r_h}$ in Fig. \ref{5d1cBH}, and the unit vector field $n$ on a portion of the $\Theta-r_h$ plane with $\tau = 3.2r_0, 2r_0$ in Fig. \ref{d51cBH}, respectively. From Fig. \ref{5d1cBH}, it is a simple matter to observe that there is one thermodynamically stable and one thermodynamically unstable black hole branch for $2.95r_0 = \tau_a < \tau < \tau_b  = 3.29r_0$, and one thermodynamically stable black hole branch for $\tau < \tau_a = 2.95r_0$.

Although in Fig.\ref{d51cBHtau32}, the zero points are located at $(r_h/r_0,\Theta) = (0.57,\pi/2)$, and $(1.27,\pi/2)$, respectively. Thus, one can read the winding numbers $w_i$ for the blue contours $C_i$: $w_1 = -1$, $w_2 = 1$, and the topological number at this inverse temperature $\tau= 3.2r_0$ is $W = -1 +1 = 0$; But in Fig. \ref{d51cBHtau2}, the zero point is located at $(r_h/r_0,\Theta) = (3.29,\pi/2)$, thus the winding number for the blue contour $C_3$ is $w_3 = 1$, so the topological number at this inverse temperature $\tau = 2r_0$ is $W = 1$. Thus, we find that the topological number is temperature dependent: it is $W = 0$ (at inverse temperature $2.95r_0 = \tau_a < \tau < \tau_b  = 3.29r_0$) or $W = 1$ (at inverse temperature $\tau < \tau_a = 2.95r_0$). At the point of the critical inversion temperature $\tau  = \tau_a$, the black hole occurs a novel temperature-dependent thermodynamic topological phase transition. The critical point corresponding to the inverse temperature $\tau_a$ should be a thermodynamic topological higher-order phase transition critical point.

\subsubsection{Critical value of electric charge parameter}

\begin{figure}[t]
\centering
\includegraphics[width=0.5\textwidth]{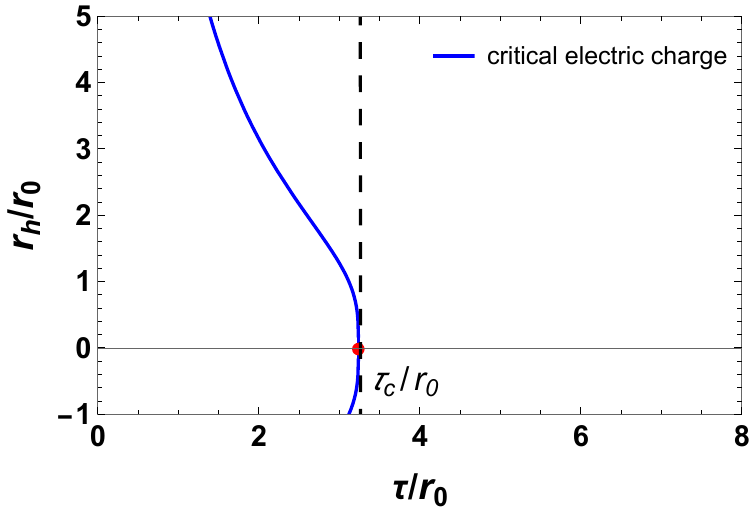}
\caption{Zero points of the vector $\phi^{r_h}$ shown in the $r_h-\tau$ plane with $q_1/r_0^2 = 15/(2\pi)$, $q_2 = q_3 = 0$, and $Pr_0^2 = 0.1$. There is one thermodynamically stable five-dimensional static charged AdS black hole in K-K gauged supergravity theory for $\tau < \tau_c  = 3.24r_0$.
\label{5dKK}}
\end{figure}

\begin{figure}[h]
\centering
\includegraphics[width=0.5\textwidth]{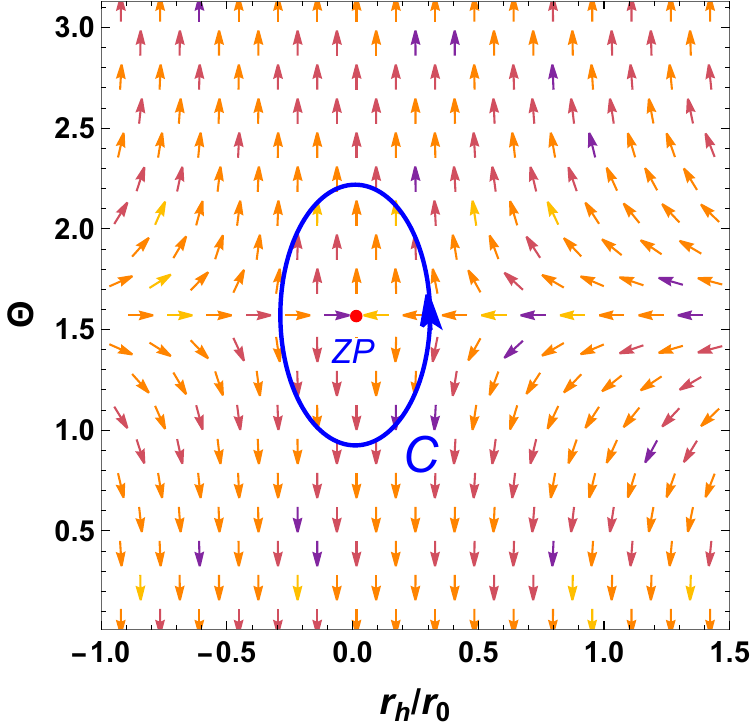}
\caption{The arrows represent the unit vector field $\hat{n}$ on a portion of the $r_h-\Theta$ plane with $q_1/r_0^2 = 15/(2\pi)$, $q_2 = q_3 = 0$, and $Pr_0^2 = 0.1$. The critical point (CP) marked with a red dot is at $(r_h/r_0, \Theta) = (0, \pi/2)$. The blue contour $C$ is a closed loop enclosing the critical point.
\label{d5KK}}
\end{figure}

In the following, we will calculate the critical value of the electric charge parameter. When the electric charge parameters $q_1 = q$ and $q_2 = q_3 = 0$, the inverse temperature $\tau$ in Eq. (\ref{tau5d}) becomes
\be
\tau = \frac{2\pi(3r_h^2 +2q)}{\sqrt{r_h^2 +q}(8\pi Pr_h^2 +4\pi Pq +3)} \, .
\ee
Using the definition of vector $\varphi$ in Eq. (\ref{varphi}), one can obtain the components of the vector $\varphi$ as
\be
\varphi^{r_h} = -\frac{2\pi r_h[4\pi P(6r_h^4 +9r_h^2q +4q^2) -9r_h^2 -12q]}{(r_h^2 +q)^{\frac{3}{2}}[4\pi P(2r_h^2 +q) +3]^2} \, , \qquad \varphi^{\Theta} = -\cot\Theta\csc\Theta \, .
\ee
Thus, as $r_h \to 0$, the critical value of the electric charge parameter $q_c$ can be determined by solving the equation $\varphi^{r_h} = 0$, which yields:
\be
q_c = \frac{3}{4\pi P} \, .
\ee
Hence, for $q \ge q_c$ , the static charged AdS black hole in five-dimensional K-K gauged supergravity exhibits the topological number of $W = 1$. However, in the case where $0 < q < q_c$, the topological number transitions from $W = 0$ (at low temperatures) to $W = 1$ (at high temperatures).

Taking $q_1/r_0^2 = q_c/r_0^2 = 15/(2\pi)$, $q_2 = q_3 = 0$, and $Pr_0^2 = 0.1$, we plot the zero points of the component $\phi^{r_h}$ in Fig. \ref{5dKK}, and the unit vector field $\hat{n}$ on a portion of the $\Theta-r_h$ plane in Fig. \ref{d5KK}, respectively. In Fig. \ref{5dKK}, one can observe that there are one thermodynamically stable five-dimensional static charged AdS black hole in K-K gauged supergravity theory for $\tau < \tau_c  = 3.24r_0$. In Fig. \ref{d5KK}, the critical point (CP) is located at $(r_h/r_0,\Theta) = (0,\pi/2)$, and the topological charge of this critical point is $\hat{W} = -1$, therefore it is a conventional critical point \cite{PRD105-104003}.

\subsection{$q_1 = q_2 \ne 0$, $q_3 = 0$ case (EMDA gauged supergravity)}

\begin{figure}[t]
\centering
\includegraphics[width=0.5\textwidth]{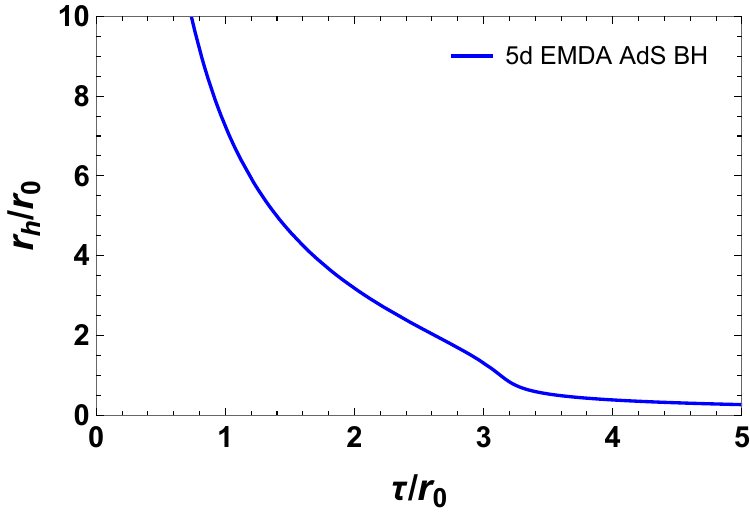}
\caption{Zero points of the vector $\phi^{r_h}$ shown in the $r_h-\tau$ plane with $q_1/r_0^2 = q_2/r_0^2 = 1$, $q_3 = 0$, and $Pr_0^2 = 0.1$. There is always one thermodynamically stable static charged AdS black hole in five-dimensional EMDA gauged supergravity theory for any value of $\tau$. Obviously, the topological number is: $W = 1$.
\label{5d2cBH}}
\end{figure}

\begin{figure}[t]
\centering
\includegraphics[width=0.5\textwidth]{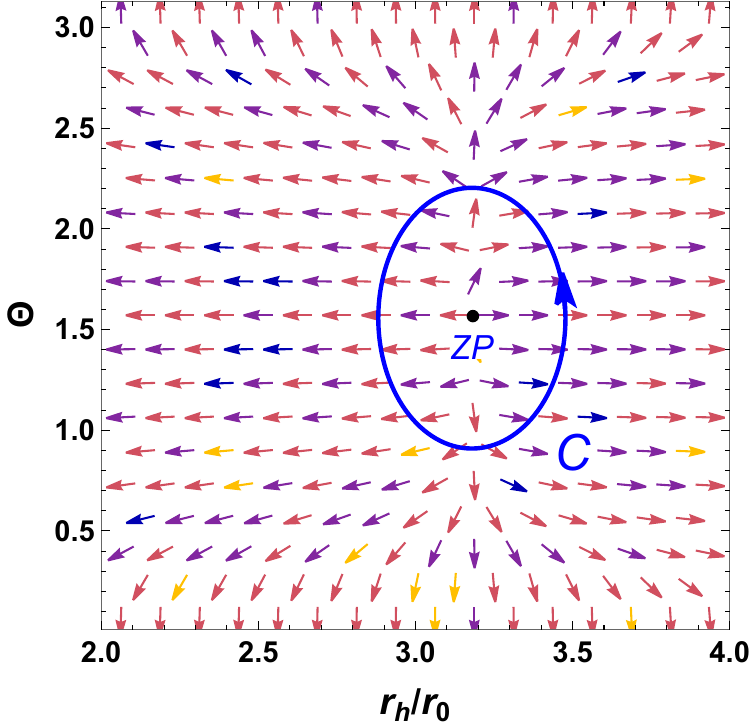}
\caption{The arrows represent the unit vector field $n$ on a portion of the $r_h-\Theta$ plane for the static charged AdS black hole in five-dimensional EMDA gauged supergravity theory with $\tau/r_0 = 2$, $q_1/r_0^2 = q_2/r_0^2 = 1$, $q_3 = 0$, and $Pr_0^2 = 0.1$. The zero point (ZP) marked with a black dot is at $(r_h/r_0, \Theta) = (3.18,\pi/2)$. The blue contour $C$ is a closed loop enclosing the zero point.
\label{d52cBH}}
\end{figure}

In this subsection, we investigate the case with $q_1 = q_2 \ne 0$, $q_3 = 0$, which represents the static charged AdS black hole in the five-dimensional EMDA gauged supergravity theory. Considering the pressure as $Pr_0^2 = 0.1$ and the electric charge parameters $q_1/r_0^2 = q_2/r_0^2 = 1$, $q_3 = 0$ for the static charged AdS black hole in five-dimensional EMDA gauged supergravity theory, we plot the zero points of $\phi^{r_h}$ in the $r_h-\tau$ plane in Fig. \ref{5d2cBH}, and the unit vector field $n$ on a portion of the $\Theta-r_h$ plane with $\tau/r_0 = 2$ in Fig. \ref{d52cBH}. Obviously, there is only one thermodynamically stable static charged AdS black hole in five-dimensional EMDA gauged supergravity theory for any value of $\tau$. In Fig. \ref{d52cBH}, one can observe that the zero point is located at $(r_h/r_0, \Theta) = (3.18,\pi/2)$. Based upon the local property of the zero point, we can easily obtain the topological number $W = 1$ for the static charged AdS black hole in five-dimensional EMDA gauged supergravity theory.

\subsection{$q_1 \ne q_2 \ne 0$, $q_3 = 0$ case ($D = 5$ AdS Horowitz-Sen solution)}

\begin{figure}[t]
\centering
\includegraphics[width=0.5\textwidth]{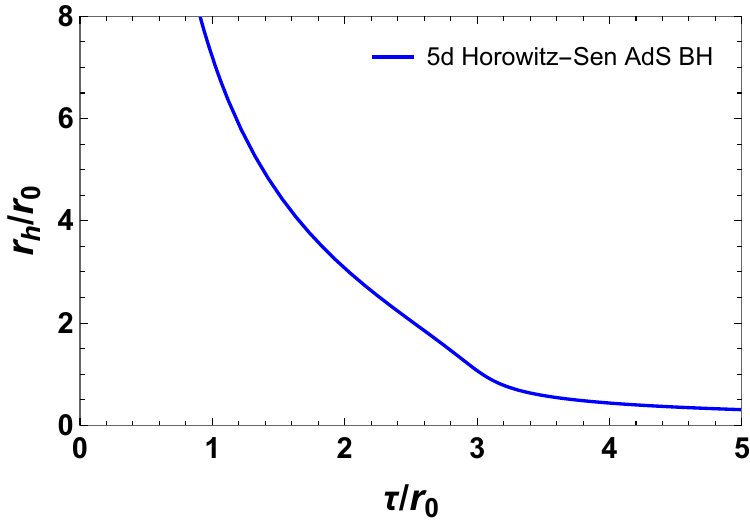}
\caption{Zero points of the vector $\phi^{r_h}$ shown in the $r_h-\tau$ plane with $q_1/r_0^2 = 1$, $q_2/r_0^2 = 2$, $q_3 = 0$, and $Pr_0^2 = 0.1$. There is always one thermodynamically stable five-dimensional static charged AdS Horowitz-Sen black hole for any value of $\tau$. Obviously, the topological number is: $W = 1$.
\label{5dHS}}
\end{figure}

\begin{figure}[t]
\centering
\includegraphics[width=0.5\textwidth]{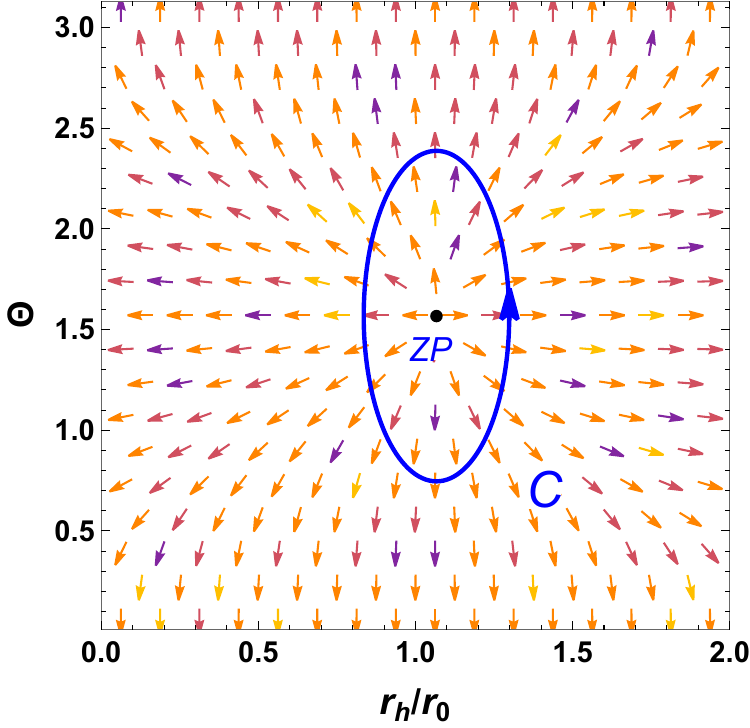}
\caption{The arrows represent the unit vector field $n$ on a portion of the $r_h-\Theta$ plane with $\tau/r_0 = 3$, $q_1/r_0^2 = 1$, $q_2/r_0^2 = 2$, $q_3 = 0$, and $Pr_0^2 = 0.1$. The zero point (ZP) marked with a black dot is at $(r_h/r_0, \Theta) = (1.07,\pi/2)$. The blue contour $C$ is a closed loop enclosing the zero point.
\label{d5HS}}
\end{figure}

In this subsection, we discuss a more general case to the last subsection, focusing on the case in which the electric charge parameters are $q_1 \ne q_2 \ne 0$ and $q_3 = 0$, which corresponds to the five-dimensional static charged AdS Horowitz-Sen black hole solution \cite{1108.5139}. Here, we would like to begin by exploring an important issue. As the smaller electric charge parameter equals zero, the five-dimensional static charged AdS Horowitz-Sen black hole reduces to the five-dimensional static charged AdS black hole in K-K gauged supergravity theory in Sec. \ref{IVA}. For the latter, the topological number $W$ is temperature-dependent; it is $W = 1$ for large electric charge parameter but can be $W = 0$ (at low temperatures) or $W = 1$ (at high temperatures) for small electric charge parameter. This raises the question: is there a critical value for the smaller electric charge parameter below which a temperature-dependent thermodynamic topological phase transition occurs?

When the electric charge parameters $q_1 \ne q_2 \ne 0$, $q_3 = 0$, the inverse temperature $\tau$ in Eq. (\ref{tau5d}) becomes
\be
\tau = \frac{2\pi[3r_h^4 +2r_h^2(q_1 +q_2) +q_1q_2]}{r_h\sqrt{r_h^2 +q_1}\sqrt{r_h^2 +q_2}[8\pi Pr_h^2 +4\pi P(q_1 +q_2) +3]} \, .
\ee
By employing the definition of the vector $\varphi$ given in Eq. (\ref{varphi}), and by solving the equation $\varphi^{r_h} = 0$, while taking the limit as $r_h \to 0$ , the critical value for the smaller electric charge parameter, $q_{1c}$, is found to be
\be
q_{1c} = 0 \, .
\ee
Therefore, the smaller electric charge parameter does not have the critical value described above, allowing the temperature-dependent thermodynamic topological phase transitions to occur.

Taking the pressure as $Pr_0^2 = 0.1$ and the electric charge parameters $q_1/r_0^2 = 1$, $q_2/r_0^2 = 2$, and $q_3 = 0$ for the five-dimensional static charged AdS Horowitz-Sen  black hole, we plot the zero points of $\phi^{r_h}$ in the $r_h-\tau$ plane in Fig. \ref{5dHS}, and the unit vector field $n$ on a portion of the $\Theta-r_h$ plane with $\tau/r_0 = 3$ in Fig. \ref{d5HS}. It is easy to observe that there is only one thermodynamically stable five-dimensional static charged AdS Horowitz-Sen black hole for any value of $\tau$. In Fig. \ref{d5HS}, one can find that the zero point is located at $(r_h/r_0, \Theta) = (1.07,\pi/2)$. Based upon the local property of the zero point, we can straightforwardly obtain the topological number $W = 1$ for five-dimensional static charged AdS Horowitz-Sen black hole.

\subsection{$q_1 = q_2 = q_3 \ne 0$ case (RN-AdS$_5$)}

\begin{figure}[t]
\centering
\includegraphics[width=0.5\textwidth]{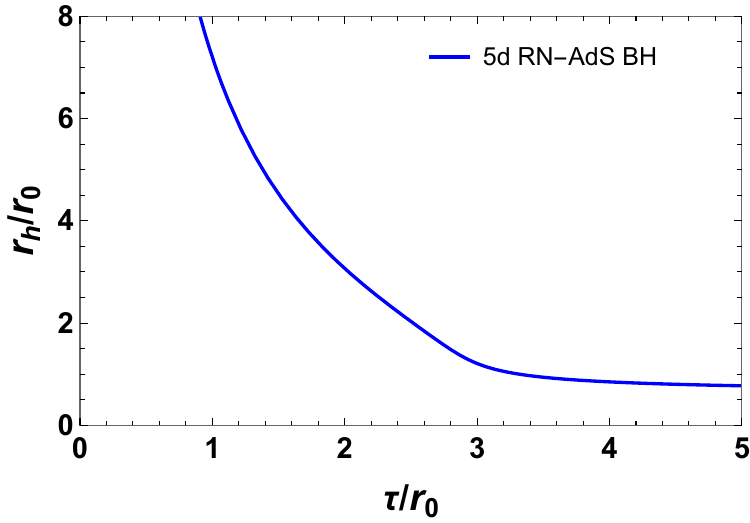}
\caption{Zero points of the vector $\phi^{r_h}$ shown in the $r_h-\tau$ plane with $q_1/r_0^2 = q_2/r_0^2 = q_3/r_0^2 = 1$, and $Pr_0^2 = 0.1$. There is one thermodynamically stable five-dimensional RN-AdS black hole for any value of $\tau$.
\label{5dRNAdS}}
\end{figure}

\begin{figure}[h]
\centering
\includegraphics[width=0.5\textwidth]{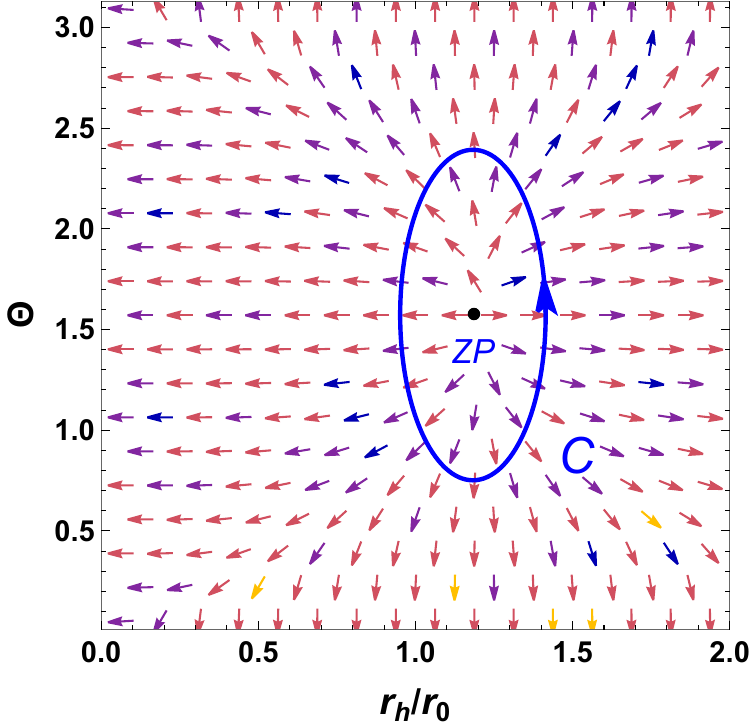}
\caption{The arrows represent the unit vector field $n$ on a portion of the $r_h-\Theta$ plane for the five-dimensional RN-AdS black hole with $\tau/r_0 = 3$, $q_1/r_0^2 = q_2/r_0^2 = q_3/r_0^2 = 1$, $Pr_0^2 = 0.1$. The zero point (ZP) marked with a black dot is at $(r_h/r_0, \Theta) = (1.21,\pi/2)$. The blue contour $C$ is a closed loop enclosing the zero point.
\label{d5RNAdS}}
\end{figure}

Considering the pressure as $Pr_0^2 = 0.1$ and the three electric charge parameters $q_1/r_0^2 = q_2/r_0^2 = q_3/r_0^2 = 1$ for the five-dimensional RN-AdS black hole, we show the zero points of $\phi^{r_h}$ in the $r_h-\tau$ plane in Fig. \ref{5dRNAdS}, and the unit vector field $n$ on a portion of the $\Theta-r_h$ plane with $\tau/r_0 = 3$ in Fig. \ref{d5RNAdS}, respectively. Based on the local property of the zero point, one can easily indicate that the topological number is: $W = 1$, which is consistent with the result given in Ref. \cite{PRL129-191101}.

\subsection{$q_1 \ne q_2 \ne q_3 \ne 0$ case (STU gauged supergravity)}\label{IVE}

\begin{figure}[t]
\centering
\includegraphics[width=0.5\textwidth]{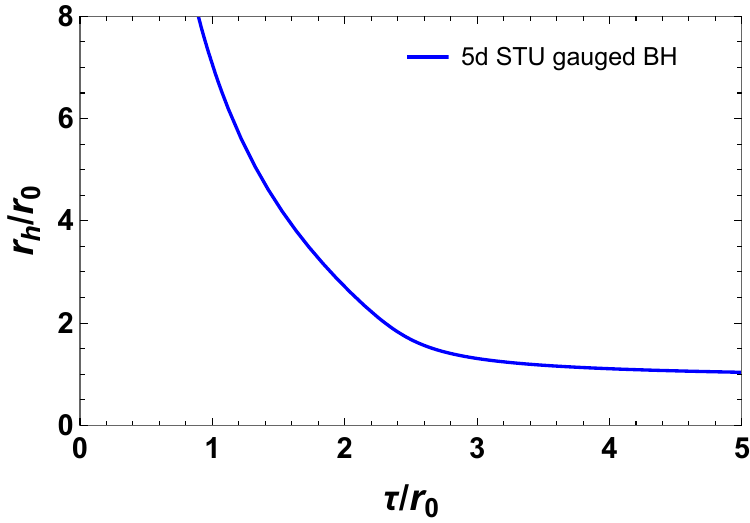}
\caption{Zero points of the vector $\phi^{r_h}$ shown in the $r_h-\tau$ plane with $q_1/r_0^2 = 1$, $q_2/r_0^2 = 2$, $q_3/r_0^2 = 3$, and $Pr_0^2 = 0.1$. There is always one thermodynamically stable five-dimensional static three-charge AdS black hole in STU gauged supergravity theory for any value of $\tau$. Obviously, the topological number is: $W = 1$.
\label{5dSTU}}
\end{figure}

\begin{figure}[h]
\centering
\includegraphics[width=0.5\textwidth]{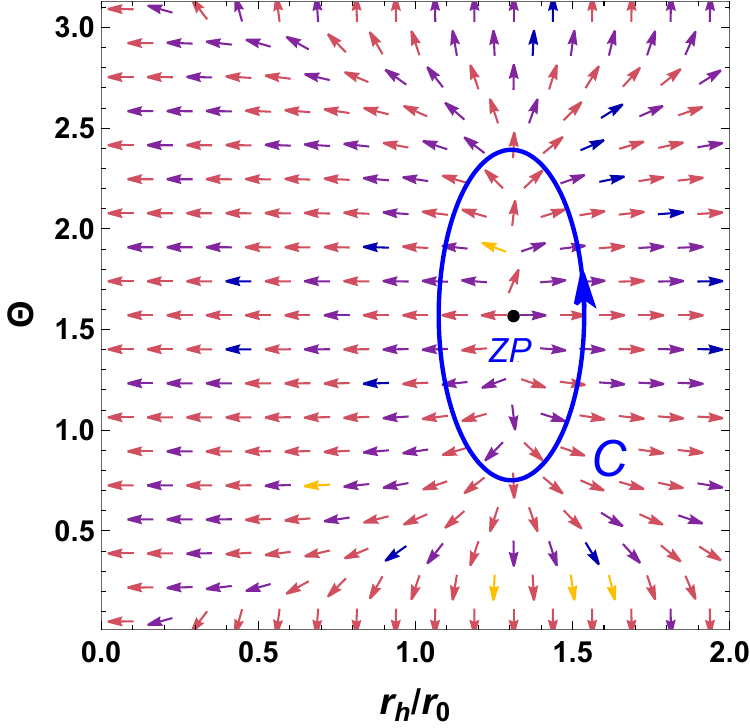}
\caption{The arrows represent the unit vector field $n$ on a portion of the $r_h-\Theta$ plane for the five-dimensional static three-charge AdS black hole in STU gauged supergravity theory with $\tau/r_0 = 3$, $q_1/r_0^2 = 1$, $q_2/r_0^2 = 2$, $q_3/r_0^2 = 3$, and $Pr_0^2 = 0.1$. The zero point (ZP) marked with a black dot is at $(r_h/r_0, \Theta) = (1.31,\pi/2)$. The blue contour $C$ is a closed loop enclosing the zero point.
\label{d5STU}}
\end{figure}

In this subsection, we investigate the most general static three-charge AdS black hole in five-dimensional STU gauged supergravity theory, namely, $q_1 \ne q_2 \ne q_3 \ne 0$ case. We take $q_1/r_0^2 = 1$, $q_2/r_0^2 = 2$, $q_3/r_0^2 = 3$, and $Pr_0^2 = 0.1$, and then plot the zero points of the component $\phi^{r_h}$ in Fig. \ref{5dSTU}, and the unit vector field $n$ on a portion of the $\Theta-r_h$ plane with $\tau/r_0 = 3$ in Fig. \ref{d5STU}, respectively. It is easy to see that there is always one thermodynamically stable five-dimensional static three-charge AdS black hole in STU gauged supergravity theory for any value of $\tau$. In Fig. \ref{d5STU}, we can observe a zero point at $(r_h/r_0, \Theta) = (1.31,\pi/2)$. Based upon the local property of the zero points, it is simple to indicate that the topological number $W = 1$ for the five-dimensional static three-charge AdS black hole in STU gauged supergravity theory.

\section{Conclusions and outlooks}\label{V}

\begin{table*}[t]
\caption{The topological number $W$, numbers of generation and annihilation points for the four-dimensional static multi-charge AdS black holes in gauged supergravity.}
\resizebox{\textwidth}{!}{
\begin{tabular}{c|c|c|c}
\hline\hline
BH solution & $W$ & Generation point &Annihilation point\\ \hline
$q_1 = q$, $q_2 = q_3 = q_4 = 0$ case & 0 & 0 & 1\\
$q_1 = q_2 = q$, $q_3 = q_4 = 0$ case ($q \ge \frac{\sqrt{6}}{4\sqrt{\pi P}}$) & 1 & 0 & 0\\
$q_1 = q_2 = q$, $q_3 = q_4 = 0$ case ($q < \frac{\sqrt{6}}{4\sqrt{\pi P}}$) & 0 (at cold temperatures) or 1 (at high temperatures) & 0 & 1\\
$q_1 \ne q_2 \ne 0$, $q_3 = q_4 = 0$ case ($q_1 \ge \frac{3}{8\pi Pq_2}$) & 1 & 0 & 0\\
$q_1 \ne q_2 \ne 0$, $q_3 = q_4 = 0$ case ($q_1 < \frac{3}{8\pi Pq_2}$) & 0 (at cold temperatures) or 1 (at high temperatures) & 0 & 1\\
$q_1 = q_2 = q_3 = q_4 \ne 0$ case (RN-AdS$_4$) & 1 & 0 & 0 \\
$q_1 = q_2 \ne 0$, $q_3 = q_4 \ne 0$ case & 1 & 0 & 0 \\
$q_1 \ne q_2 \ne q_3 \ne q_4 \ne 0$ case & 1 & 0 & 0 \\
\hline\hline
\end{tabular}}
\label{TableI}
\end{table*}

\begin{table*}[h]
\caption{The topological number $W$, numbers of generation and annihilation points for the five-dimensional static multi-charge AdS black holes in gauged supergravity.}
\resizebox{\textwidth}{!}{
\begin{tabular}{c|c|c|c}
\hline\hline
BH solution & $W$ & Generation point &Annihilation point\\ \hline
$q_1 = q$, $q_2 = q_3 = 0$ case ($q \ge \frac{3}{4\pi P}$) & 1 & 0 & 0\\
$q_1 = q$, $q_2 = q_3 = 0$ case ($q < \frac{3}{4\pi P}$) & 0 (at cold temperatures) or 1 (at high temperatures) & 0 & 1\\
$q_1 = q_2 \ne 0$, $q_3 = 0$ case & 1 & 0 & 0\\
$q_1 = q_2 = q_3 \ne 0$ case (RN-AdS$_5$) & 1 & 0 & 0 \\
$q_1 \ne q_2 \ne q_3 \ne 0$ case & 1 & 0 & 0 \\
\hline\hline
\end{tabular}}
\label{TableII}
\end{table*}

In this paper, making use of the generalized off-shell Helmholtz free energy, we investigate the topological number of the four-dimensional static multi-charge AdS black holes in gauged supergravity theory \cite{NPB554-237} and the five-dimensional static multi-charge AdS black holes in gauged supergravity theory \cite{NPB553-317}. In gauged supergravity theory, four- and five-dimensional static charged AdS black holes have four and three independent electric charge parameters, respectively. In this study, we investigate the effect of the electric charge parameter configurations in static charged AdS black holes on the thermodynamic topological classification in the context of four- and five-dimensional gauged supergravity theories. For each black hole case, we examine various electric charge parameter configurations corresponding to several well-known truncated supergravity solutions and determine their topological numbers, respectively. The findings are summarized in Tables \ref{TableI}-\ref{TableII}. We find that the topological number of the static charged AdS black holes in four-dimensional K-K gauged supergravity theory is $W = 0$, while that of the static charged AdS black holes in four-dimensional gauged $-iX^0X^1$-supergravity and STU gauged supergravity theories, and five-dimensional EMDA gauged supergravity and STU gauged supergravity, as well as five-dimensional static charged AdS Horowitz-Sen black hole are both $W = 1$.

Furthermore, we observe a novel temperature-dependent thermodynamic topological phase transition that can happen in the four-dimensional static charged AdS black hole in EMDA gauged supergravity theory, the four-dimensional static charged AdS Horowitz-Sen black hole, and the five-dimensional static charged AdS black hole in K-K gauged supergravity theory. In other words, in our analysis of four-dimensional black hole cases, we demonstrate that when only two electric charge parameters are nonzero (with the other two set to zero), the thermodynamic topological number $W$ exhibits a temperature-dependent behavior. Specifically, $W = 1$ when the two electric charge parameters are large, whereas for smaller electric charge parameters, $W$ can either be $W = 0$ at lower temperatures or $W = 1$ at higher temperatures. This behavior is mirrored in the context of five-dimensional black hole cases within the framework of K-K gauged supergravity theory, where the static charged AdS black hole also displays a topological number ($W = 1$) for the large electric charge parameter. However, for the smaller electric charge parameter, the topological number $W$ once again shows a temperature dependence: $W = 0$ at cold temperatures or $W = 1$ at high temperatures.

Therefore, we believe that the current studies related to the thermodynamic topological classes of black holes are still only the tip of the iceberg, and it is worthwhile to explore the nature of the topological number of black hole thermodynamics more deeply. A most related issue is to explore whether there are other black hole solutions in gauged supergravity theories that can also happen this novel temperature-dependent thermodynamic topological phase transition. As mentioned above, we only investigated the topological numbers of static charged AdS black holes in several famous four- and five-dimensional truncated supergravity models, and the static charged AdS black hole solutions in other supergravity theories can be investigated in the future, e.g., the $S^3$-supergravity model \cite{1108.5139,NPB250-385,NPB459-125}, the $ST^2$-supergravity model \cite{PRD87-044055}, etc.

\acknowledgments
We are greatly indebted to the anonymous referee for the constructive comments to improve the presentation of this work. We also
thank Professor Shuang-Qing Wu for helpful suggestions and detailed discussions. This work is supported by the National Natural Science
Foundation of China (NSFC) under Grants No. 12205243 and No. 12375053, by the Sichuan Science and Technology Program under Grant No. 2023NSFSC1347,
by the Sichuan Youth Science and Technology Innovation Research Team under Grant No. 21CXTD0038, and by the Doctoral Research Initiation
Project of China West Normal University under Grant No. 21E028.

\end{document}